\documentclass{IEEEtaes}

\usepackage[colorlinks,urlcolor=blue,linkcolor=blue,citecolor=blue]{hyperref}

\usepackage{color,array,amsthm}
\usepackage{makecell}
\usepackage{graphicx,lineno}

\usepackage{multirow,booktabs,tabularx,longtable,tabu,threeparttable}

\usepackage{amsbsy,amsmath,amsfonts,cases,bm,cite,subfigure,gensymb,amssymb}
\usepackage[ruled,vlined]{algorithm2e}

\usepackage{scalerel}
\usepackage{tikz}

\usetikzlibrary{svg.path}

\definecolor{orcidlogocol}{HTML}{A6CE39}
\tikzset{
	orcidlogo/.pic={
		\fill[orcidlogocol] svg{M256,128c0,70.7-57.3,128-128,128C57.3,256,0,198.7,0,128C0,57.3,57.3,0,128,0C198.7,0,256,57.3,256,128z};
		\fill[white] svg{M86.3,186.2H70.9V79.1h15.4v48.4V186.2z}
		svg{M108.9,79.1h41.6c39.6,0,57,28.3,57,53.6c0,27.5-21.5,53.6-56.8,53.6h-41.8V79.1z M124.3,172.4h24.5c34.9,0,42.9-26.5,42.9-39.7c0-21.5-13.7-39.7-43.7-39.7h-23.7V172.4z}
		svg{M88.7,56.8c0,5.5-4.5,10.1-10.1,10.1c-5.6,0-10.1-4.6-10.1-10.1c0-5.6,4.5-10.1,10.1-10.1C84.2,46.7,88.7,51.3,88.7,56.8z};
	}
}

\newcommand\orcidicon[1]{\href{https://orcid.org/#1}{\mbox{\scalerel*{
				\begin{tikzpicture}[yscale=-1,transform shape]
				\pic{orcidlogo};
				\end{tikzpicture}
			}{|}}}}

\usepackage{hyperref} 

\AtBeginDocument{\hypersetup{pdfborder={0 0 0}}}

\begin{document}

\title{Globally Optimized TDOA High Frequency Source Localization Based on Quasi-Parabolic Ionosphere Modeling and Collaborative Gradient Projection}

\author{Wenxin Xiong}
\member{Graduate Student Member, IEEE}
\affil{Department of Computer Science, University of Freiburg, Freiburg 79110, Germany} 

\author{Christian Schindelhauer}
\member{Member, IEEE}
\affil{Department of Computer Science, University of Freiburg, Freiburg 79110, Germany} 

\author{Hing Cheung So}
\member{Fellow, IEEE}
\affil{Department of Electrical Engineering, City University of Hong Kong, Hong Kong, China}


\receiveddate{This work was supported by the state graduate funding coordinated by Uni Freiburg.}

\corresp{{\itshape (Corresponding author: Wenxin Xiong)}.}

\authoraddress{Wenxin~Xiong and Christian~Schindelhauer are with the Department of Computer Science, University of Freiburg, Freiburg 79110, Germany (e-mail: w.x.xiong@outlook.com; schindel@informatik.uni-freiburg.de). \\Hing~Cheung~So is with the Department of Electrical Engineering, City University of Hong Kong, Hong Kong, China (e-mail: hcso@ee.cityu.edu.hk).}

\supplementary{This is the accepted version. The final version of this paper has been published in the IEEE Transactions on Aerospace and Electronic Systems: W. Xiong, C. Schindelhauer and H. C. So, "Globally Optimized TDOA High-Frequency Source Localization Based on Quasi-Parabolic Ionosphere Modeling and Collaborative Gradient Projection," in IEEE Transactions on Aerospace and Electronic Systems, vol. 59, no. 1, pp. 580-590, Feb. 2023, doi: 10.1109/TAES.2022.3185971. The copyright is with IEEE.}

\markboth{XIONG ET AL.}{GLOBALLY OPTIMIZED TDOA HF LOCALIZATION VIA COLLABORATIVE GRADIENT PROJECTION}
\maketitle

\begin{abstract}
We investigate the problem of high frequency (HF) source localization using the time-difference-of-arrival (TDOA) observations of ionosphere-refracted radio rays based on quasi-parabolic (QP) modeling. An unresolved but pertinent issue in such a field is that the existing gradient-type scheme can easily get trapped in local optima for practical use. This will lead to the difficulty in initializing the algorithm and finally degraded positioning performance if the starting point is inappropriately selected. In this paper, we develop a collaborative gradient projection (GP) algorithm in order to globally solve the highly nonconvex QP-based TDOA HF localization problem. The metaheuristic of particle swarm optimization (PSO) is exploited for information sharing among multiple GP models, each of which is guaranteed to work out a critical point solution to the simplified maximum likelihood formulation. Random mutations are incorporated to avoid the early convergence of PSO. Rather than treating the geolocation of HF transmitter as a pure optimization problem, we further provide workarounds for addressing the possible impairments and challenges when the proposed technique is applied in practice. Numerical results demonstrate the effectiveness of our PSO-assisted re-initialization strategy in achieving the global optimality, and the superiority of our method over its competitor in terms of positioning accuracy.
\end{abstract}


\section{INTRODUCTION}
H{\scshape igh} frequency (HF) radio signals reflected/refracted from the ionosphere based on skywave propagation are increasingly exploited for communication beyond the horizon at distances in the order of several thousands of kilometers \cite{AJain1,TWang,SHuang,AJain2,AJain3}. In a nutshell, solar radiation ionizes the ionosphere, thus producing free electrons that can influence the propagation of HF radio signals \cite{AJain3}. Due to the presence of distinct ionospheric layers in which the degree of ionization exhibits discrepancies, particular ionospheric models should be considered to derive analytic expressions describing the radio-ray trajectories therethrough \cite{DBilitza,AHDeVoogt,TACroft}.

A simple but effective modeling option for ray parameter calculations that takes the ionosphere medium into account is the monolayer quasi-parabolic (QP) model, defined by a parabola-like equation in electron density versus height. Originally reported during the 1950s \cite{AHDeVoogt} and 60s \cite{TACroft}, QP has been a time-tested means of ionospheric radio propagation analysis \cite{CHou}. Typically, it has been successfully utilized for HF source localization based on the sensor-collected time-difference-of-arrival (TDOA) observations for ionosphere-refracted radio rays \cite{AJain1,TWang,SHuang}. The authors of \cite{AJain1} discuss the limitations of the QP model and reuse a closed-form formula in \cite{EGBakhoum} for positioning. Their experimental results are disclosed in \cite{AJain3}, showing a relative localization error of 0.1--10\% of the true ground distance. Nevertheless, their focus is not on the parabola-like description but on the specular-reflection and flat-Earth approximations. In \cite{TWang}, the authors present a maximum likelihood (ML) estimator for TDOA HF source localization under the QP ionosphere model. A three-step heuristic procedure is therewith outlined. However, this approach might succumb to the ambiguity associated with ray takeoff angle variables and thus be prone to failures. The simulation studies in \cite{SHuang} have demonstrated that the relative geolocation error (RGE) of the heuristic can be as large as 50\% of the ground distance in a normal noise level if it is randomly initialized. For overcoming these drawbacks, the authors of \cite{SHuang} analyze the geometry of feasible region for the ML problem, and then put forward a generalized projected gradient descent (GPGD) method to find the location estimate at a critical point of it.

The excellent work of \cite{SHuang} still leaves room for improvement, as the performance of GPGD depends to a large extent on the initialization strategies. In their numerical investigations \cite{SHuangCode}, the authors of \cite{SHuang} resort to a threshold for the difference between the source location and its estimate, to exclude solution points that are unlikely to be the global ones in the simulation protocol. However, such a criterion is impracticable in real-world localization scenarios since the true source position is unknown to the algorithm at the problem-solving stage. Without specifying a starting point close enough to the global optimum, the gradient-type solver relied on to address the highly nonconvex optimization problem may suffer from significant performance degradation. Taking our simulation results in Table \ref{table_RMSE} as an example, the RGE of the randomly initialized GPGD method is about 28.46\% of the ground distance, whereas this will be reduced to 0.31\% if the unrealistic threshold is imposed. Clearly, the concerns about initialization limit the applicability of GPGD to onsite implementation.

Recent advances in the hybrid stochastic and deterministic techniques have in a way ironed out the difficulty in solving intricate optimization problems, e.g., those involved with the nonlinearity, nonconvexity, and/or discontinuity \cite{ZYan}. In this paper, specifically, we are motivated by the successful applications of stochastically coordinated gradient-type local minimizers in achieving the global optimality under nonconvexity \cite{ZYan}. As our main contribution herein, we devise a globally optimal collaborative gradient projection (GP) scheme for HF source localization using the TDOA measurements of QP-modeled ionosphere-refracted radio rays. We adopt the metaheuristic of particle swarm optimization (PSO) for information exchange among multiple GP models, with each of the individuals being guaranteed to converge to a critical point of the ML estimation problem. We further introduce the procedure of random mutation into the collaborative GP (CGP) framework, for avoiding the prematurity of PSO in the early stages \cite{PSAndrews}. The convergence properties are also analyzed in detail. Ultimately, we conduct numerical studies to demonstrate the positioning accuracy superiority of the proposed method over the state-of-the-art solution in \cite{SHuang}. From again Table \ref{table_RMSE}, our CGP algorithm leads to a localization error much smaller than the GPGD counterparts, amounting to a relative error of merely 0.31\textperthousand~in the corresponding scenario.

The remainder of the paper is organized as follows. Section \ref{Sect_PF} describes the QP model and the HF source localization formulation to be handled. Our CGP algorithm is presented in Section \ref{Sect_AD}. Section \ref{Sect_NR} includes the performance evaluations, and necessary discussions of the related practical issues. Finally, Section \ref{Sect_Cc} draws the concluding remarks.

\textit{Notations:} Vectors and matrices are denoted by the lowercase bold and uppercase bold letters, respectively. For better visibility, Tables \ref{tb:notation} and \ref{tb:variable} list the main symbols used in this paper and describe the important variables beforehand, respectively.

\begin{table}[!htbp]
\begin{center}
\caption{Notations}
\begin{tabular}{c|l}
	\toprule
	\multicolumn{1}{m{1cm}}{\centering \bfseries Symbol}
	&\multicolumn{1}{m{6.5cm}}{\centering \bfseries Definition}\\
	\midrule
    ${\|\cdot\|}_2$ & $\ell_2$-norm. \\
    $\bm{I}_{a \times b}$ & $a \times b$ identity matrix. $\bm{I}_{a \times b} \in \mathbb{R}^{a \times b}$. \\
    $\nabla_{\bm{x}}(\cdot)$ & Gradient of a function at $\bm{x}$. $\nabla_{\bm{x}}(\cdot) \in \mathbb{R}^{3}$. \\
    $\Pi_{\Omega}(\bm{z})$ & Projection of a point $\bm{z}$ onto a set $\Omega$. \\
    $\rho(\cdot)$ & Schedule to adjust $\tau$ in GP for convergence speedup. \\
    $\leqq$ & Vector inequality (element-wise version of $\leq$). \\
    $[\cdot]_{i}$ & The $i$th element of a vector. \\
    $\{ \cdot \}$ & Either a sequence or a set, depending on the context. \\
	$(\cdot)^{(k)}$ & Iteration index of GP or PSO, context-dependent. \\
	$(\cdot)_{(j)}$ & Iteration index of AP. \\
	$\mathcal{M}(\cdot)$ & Mutation operator in PSO. \\
	\texttt{rand} & Uniformly distributed random number. \texttt{rand} $\in (0,1)$. \\
	$\mathcal{B}$ & Borel subsets of $\mathbb{R}^{N}$. \\
	$v(\cdot)$ & Lebesgue measure, a.k.a. the $N$-dimensional volume. \\
    \bottomrule
\end{tabular}\label{tb:notation}
\end{center}
\end{table}

\begin{table}[!htbp]
\begin{center}
\caption{Variable descriptions}
\begin{tabular}{c|l}
	\toprule
	\multicolumn{1}{m{1cm}}{\centering \bfseries Variable}
	&\multicolumn{1}{m{6.5cm}}{\centering \bfseries Definition}\\
	\midrule
    $\bm{x}$ & True source location. $\bm{x} \in \mathbb{R}^{3}$. \\
	$\tilde{\bm{x}}$ & Estimate of source location. $\tilde{\bm{x}} \in \mathbb{R}^{3}$. \\
	\multirow{2}{*}{$\tilde{\bm{x}}^{\{i\}}$} & Estimate of source location associated with \\ 
	& the $i$th MC sample. $\tilde{\bm{x}}^{\{i\}} \in \mathbb{R}^{3}$. \\
	$\bm{x}_i$ & Known position of the $i$th sensor. $\bm{x}_i \in \mathbb{R}^{3}$. \\
	\multirow{2}{*}{$D_i$} & True distance between the source and the $i$th sensor \\
	& along Earth's surface, a.k.a. the ground distance. \\
	\multirow{3}{*}{$P^{\prime}_i$} & True value of the time-of-arrival based range \\
	& measurement at the $i$th sensor when considering \\
	& the ionospheric medium, a.k.a. the group path. \\
	\multirow{2}{*}{$\bm{P}^{\prime}$} & Vector that holds all the group paths, \\
	& i.e., $\bm{P}^{\prime} = \left[ P^{\prime}_1,...,P^{\prime}_L \right]^T \in \mathbb{R}^{L}$. \\
	\multirow{2}{*}{$r_0$} & Earth radius under the spherical approximation. \\ 
	& $r_0 \approx 6371$ km. \\
	$r_b$ & Geocentric height of QP ionospheric layer base. \\
	$r_m$ & Geocentric height of maximum electron density. \\
	$y_m$ & Layer semithickness, namely, $y_m = r_m - r_b$. \\
	\multirow{2}{*}{$\gamma_i$} & Angle of radio ray associated with the $i$th sensor \\ 
	& at the bottom of ionosphere. \\
	\multirow{2}{*}{$\beta_i$} & Ray takeoff angle relative to the optical horizon \\ 
	& associated with the $i$th sensor. $\beta_i \in [0,\pi/2]$. \\
	\multirow{2}{*}{$\bm{\beta}$} & Vector that holds all the ray takeoff angles, \\ 
	& i.e., $\bm{\beta} = [\beta_1,...,\beta_L]^T \in \mathbb{R}^{L}$. \\
	$\beta^{\textup{U}}$ & Limit angle associated with the skip distance. \\ 
	$f$ & Operating frequency. \\
	$f_c$ & Critical frequency. \\
	$r_{i,1}$ & TDOA-based RD measurement at the $i$th sensor. \\
	\multirow{2}{*}{$\bm{r}$} & Vector that holds all the RD measurements, \\
	& i.e., $\bm{r} = \left[ r_{2,1},...,r_{L,1} \right]^T \in \mathbb{R}^{L-1}$. \\
	\multirow{2}{*}{$n_i$} & Zero-mean Gaussian disturbance with variance $\sigma_i^2$,\\ 
	& associated with the $i$th sensor.\\
	$\bm{\Sigma}$ & Covariance matrix for $\bm{r}$. $\bm{\Sigma} \in \mathbb{R}^{(L-1) \times (L-1)}$. \\
	$\bm{E}$ & $\bm{E} = \left[ -\bm{1}_{L-1}, \bm{I}_{(L-1)\times (L-1)} \right] \in \mathbb{R}^{(L-1) \times L}$. \\
	$\tau$ & Positive step size of the GP algorithm. \\
	$N_{\max}^{\prime}$ & Predefined maximum number of GP iterations. \\
	$N_{\textup{GP}}$ & Number of iterations for GP to converge. \\
	$N_{\textup{Quad}}$ & Number of iterations for quadratic programming. \\
	$N_{\max}$ & Predefined maximum number of PSO iterations. \\
	$N$ & Dimension of PSO search space. \\
	$N_{\textup{MC}}$ & Total of MC samples. \\
	$w_{\max}$ & Predefined upper limit of weighting factor in PSO. \\
	$w_{\min}$ & Predefined lower limit of weighting factor in PSO. \\
	$c_{1}$ & Cognitive parameter in PSO. \\
	$c_{2}$ & Social parameter in PSO. \\
	$N_{\textup{Ptcl}}$ & Number of particles used in PSO. \\
	$\gamma_{\textup{Div}}$ & Predefined tolerance for the swarm diversity. \\
    \bottomrule
\end{tabular}\label{tb:variable}
\end{center}
\end{table}

\section{PRELIMINARIES AND PROBLEM FORMULATION}

\label{Sect_PF}
\begin{figure}[!t]
	\centering
	\includegraphics[width=3.2in]{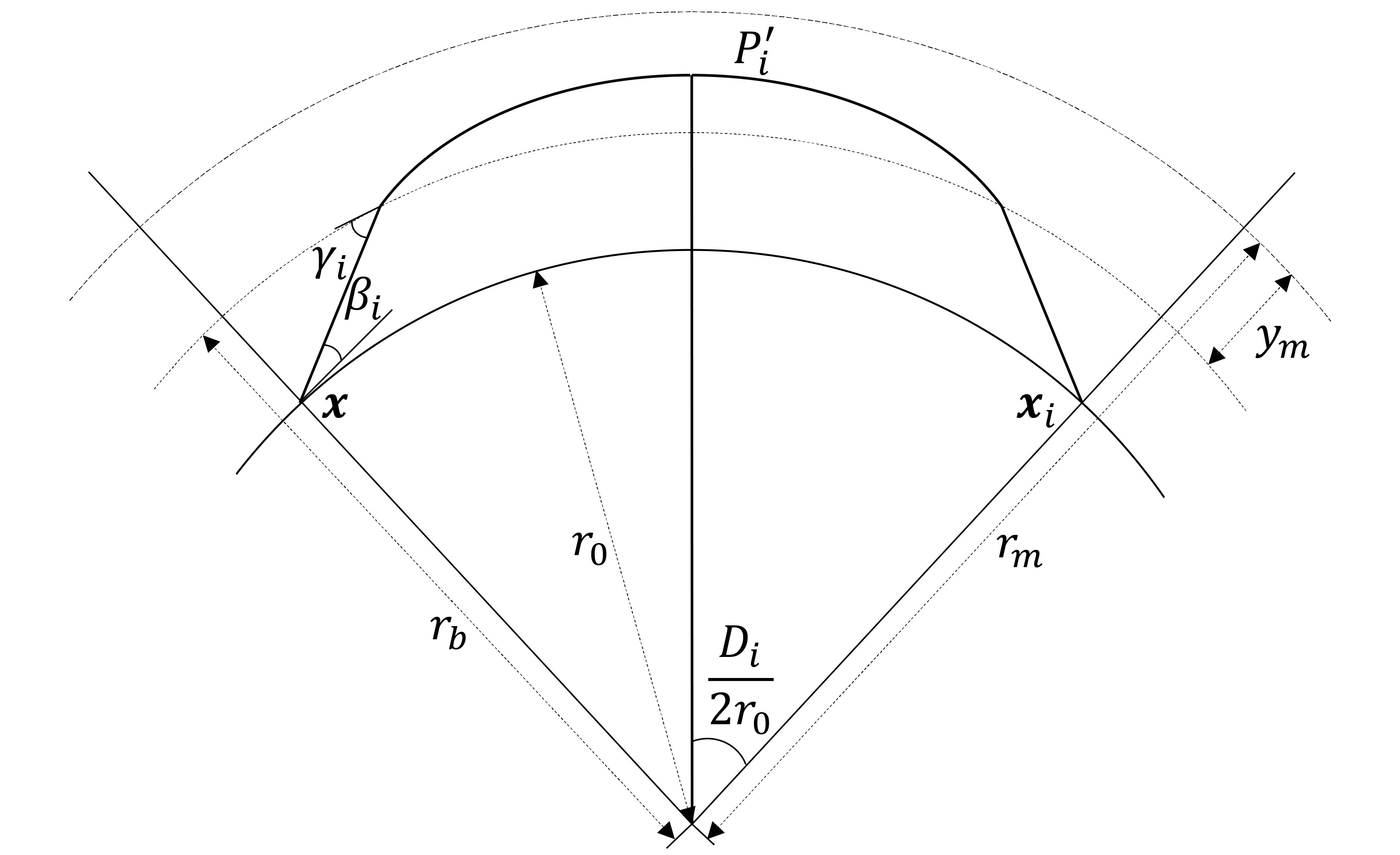}
	\caption{Illustration of radio-ray path.} 
	\label{FIG1}
\end{figure}
Our HF localization scenario comprises $L \geq 4$ coordinated sensors and a single source deployed on the ground. The known position of the $i$th sensor and unknown position of the source, both in the three-dimensional geocentric coordinate system, are denoted by $\bm{x}_i \in \mathbb{R}^3$ (for $i = 1,...,L$) and $\bm{x} \in \mathbb{R}^3$, respectively. The source emits an HF radio signal, which travels through the ionosphere and is finally collected at different sensors. Fig. \ref{FIG1} illustrates the radio-ray trajectory from the HF source via ionosphere to the $i$th sensor. Taking into account the effects of Earth curvature and utilizing the QP model \cite{TACroft}, we are able to describe the source-sensor geometry and derive exact expressions for the radio-ray trajectories as
\begin{align}{\label{QPmdl}}
&D_i\!=\!2r_0\!\left(\!(\gamma_i - \beta_i) - \tfrac{r_0 \cos \beta_i}{2 \sqrt{C_i}} \ln \tfrac{B^2 - 4AC_i}{4C_i\left( \sin \gamma_i + \tfrac{\sqrt{C_i}}{r_b} + \tfrac{B}{2 \sqrt{C_i}} \right)^2}\!\right)\!,\nonumber\\
&P^{\prime}_i = 2\bigg( r_b \sin \gamma_i - r_0 \sin \beta_i \nonumber\\
&~~+ \tfrac{1}{A}\!\left( -r_b \sin \gamma_i - \tfrac{B}{4 \sqrt{A}} \ln \tfrac{B^2 - 4AC_i}{\left( 2A r_b + B + 2 r_b \sqrt{A} \sin \gamma_i \right)^2} \right)\!\bigg),
\end{align}
where $A = 1 - \tfrac{1}{F^2} + \left( \tfrac{r_b}{F {y_m}} \right)^2$, $B = - \tfrac{2 r_m r_b^2}{F^2 y_m^2}$, and $C_i = \left(\tfrac{r_b r_m}{F {y_m}}\right)^2 - r_0^2\cos^2 \beta_i$. Readers may refer to Table \ref{tb:variable} for the detailed definitions of related variables. Note that in order to offer long-range coverage in the majority of practical situations, the operating frequency is usually selected on the premise that the ratio of it to the critical frequency, $F = f/f_c$, is larger than 1 \cite{GSSales}. In addition, we have $r_b \cos \gamma_i = r_0 \cos \beta_i$ by Snell's law, hence $D_i$ and $P^{\prime}_i$ are both treated as functions of $\bm{\beta}$.

Designating the first sensor as the reference, the TDOA-based range-difference (RD) measurement at the $i$th sensor is modeled as 
\begin{equation}{\label{TDOAmdl}}
	r_{i,1} = P^{\prime}_i - P^{\prime}_1 + n_i - n_1,~i = 2,...,L,
\end{equation}
where $n_{i}$ (for $i = 1,...,L$) denotes the uncorrelated additive noise associated with the $i$th sensor, assumed to be a zero-mean Gaussian process with variance $\sigma_{i}^2$. In a typical manner, the difference in propagation duration between a single pair of synchronized sensors is obtained by cross-correlating the received signals. Despite data transmission restrictions in such a scenario and ubiquitous error-causing environmental factors like the multipath propagation and interferences, recent years have witnessed a growing number of successful TDOA receiver networks, operational for long-range HF geolocation. For example, Jain \textit{et al.} elaborate in \cite{AJain4} on their cross-channel sounding methodology for accessing the channel impulse response and evaluating the time delays. Experimental results of the corresponding prototype system can be found in \cite{AJain3}. Another mentionable work is perhaps \cite{NXia}. Similarly, the TDOAs are produced by maximizing the cross-correlation function. What differentiates it from \cite{AJain3} is that the authors of \cite{NXia} turn instead to Kalman particle filtering for the state estimation of the HF source location and virtual heights of the ionosphere. Interested readers are referred to \cite{NXia2} for a performance comparison between \cite{AJain3} and \cite{NXia}.

The TDOA HF localization here refers to determining $\bm{x}$ using $\bm{r} = \left[ r_{2,1},...,r_{L,1} \right]^T \in \mathbb{R}^{L-1}$ and the \textit{a priori} known $\{ \bm{x}_i | i = 1,...,L \}$. Bearing an indirect relation to $r_{i,1}$ through (\ref{QPmdl}), (\ref{TDOAmdl}), and the spherical distance constraints between the source and sensors:
\begin{equation}{\label{2ndeq_cons}}
	2r_0 \sin \left( \tfrac{D_i}{2r_0} \right) = {\|\bm{x}_i - \bm{x}\|}_2,~~i = 1,...,L,
\end{equation}
$\bm{x}$ can be estimated based on the ML principle, by solving
\begin{align}{\label{ML_est}}
\min_{\bm{x},\bm{\beta}}~&(\bm{r} - \bm{E} \bm{P}^{\prime})^T \bm{\Sigma}^{-1} (\bm{r} - \bm{E} \bm{P}^{\prime}) \nonumber\\
\textup{s.t.}&~\textup{(\ref{2ndeq_cons})},~r_0 = {\|\bm{x}\|}_2,
\end{align}
as has been previously formulated in \cite{TWang}. Here, (\ref{2ndeq_cons}) relates $\bm{x}$ to $\bm{\beta}$, $\bm{P}^{\prime} = \left[ P^{\prime}_1,...,P^{\prime}_L \right]^T \in \mathbb{R}^{L}$, $\bm{\Sigma} \in \mathbb{R}^{(L-1) \times (L-1)}$ represents the covariance matrix for $\bm{r}$, and $r_0 = {\|\bm{x}\|}_2$ is the so-called Earth constraint to keep the source on the ground.

\begin{figure}[!t]
	\centering
	\includegraphics[width=3.2in]{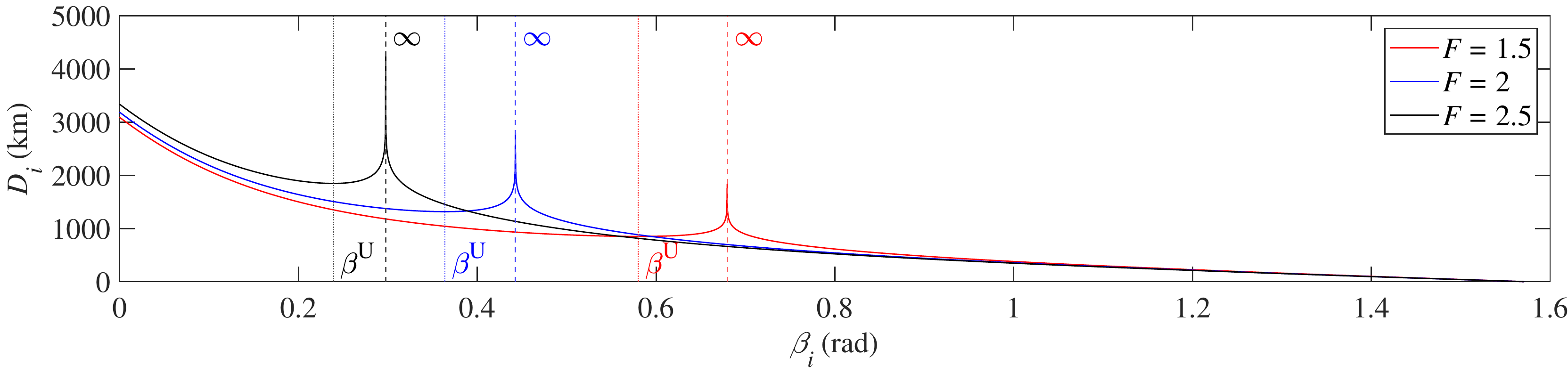}
	\caption{QP-modeled relationship between $D_i$ and $\beta_i$ when $F > 1$.} 
	\label{FIG2}
\end{figure}

As pointed out by \cite{SHuang}, certain angle constraints should be imposed on $\beta_i$ so as to be of practical significance. For illustrative purposes, Fig. \ref{FIG2} plots $D_i$ versus $\beta_i \in \left[ 0, \pi/2 \right]$ in the same system parameter settings as those for the first experiment in Section \ref{Sect_NR}, except that we change $f$ into 15, 20, and 25 MHz in the cases of $F =$ 1.5, 2, and 2.5, respectively. As the ray takeoff angle $\beta_i$ increases from 0 to an easily determinable local minimum point $\beta^{\textup{U}}$, the ground distance $D_i$ monotonically decreases, reaching the minimum range for a detectable HF source (associated with $\beta^{\textup{U}}$ and also known as (a.k.a.) the skip distance \cite{GSSales}). Once $\beta_i$ is larger than $\beta^{\textup{U}}$, $D_i$ turns to increase as it increases, until the rays eventually penetrate the ionospheric layer (corresponding to the infinities in Fig. \ref{FIG2}, whose locations can be found by solving $B^2 = 4AC_i$). One may observe here that for any $D_i$ larger than the skip distance, the ray calculation leads to two takeoff angle solutions, namely, one less than $\beta^{\textup{U}}$ and the other falling into the interval between $\beta^{\textup{U}}$ and the threshold angle for which all rays penetrate. Nonetheless, normally only the former low-angle ray will be involved in the operations of the localization system. The latter, on the other hand, will be regarded as a nuisance because it is comparatively weaker \cite{GSSales}. 

Incorporating $0 \leq \beta_i \leq \beta^{\textup{U}}$ (for $i = 1,...,L$) into the formulation, (\ref{ML_est}) is simplified as \cite{SHuang}
\begin{align}{\label{ML_est_ref}}
\min_{\bm{x} \in \Omega}g(\bm{x})\!:=\!(\bm{r} - \bm{E} \bm{P}^{\prime} (\bm{\beta}(\bm{x})))^T \bm{\Sigma}^{-1} (\bm{r} - \bm{E} \bm{P}^{\prime} (\bm{\beta}(\bm{x}))),
\end{align}
where $\bm{\beta}(\cdot)$ corresponds to a function of $\bm{x}$ in accordance with (\ref{2ndeq_cons}), $\Omega = \mathcal{S} \cap \mathcal{T}$, $\mathcal{S} = \big\{ \bm{x} \in \mathbb{R}^{3} \big| r_0 = {\|\bm{x}\|}_2 \big\}$ is a sphere defined by $r_0 = {\|\bm{x}\|}_2$, and
\begin{align}{\label{PolytopeT}}
	\mathcal{T} &= \big\{ \bm{x} \in \mathbb{R}^{3} \big| r_0^2 - 2r_0^2 \sin^2 \big( \tfrac{D_i(0)}{2r_0} \big) \leq \bm{x}_i^T \bm{x} \nonumber\\
	&~~\leq r_0^2 - 2r_0^2 \sin^2 \big( \tfrac{D_i(\beta^{\textup{U}})}{2r_0} \big),~i = 1,...,L \big\}
\end{align}
is a polytope defined by $0 \leq \beta_i \leq \beta^{\textup{U}}$ and (\ref{2ndeq_cons}), based on the assumption that the HF source and sensors are all deployed on the ground (viz., ${\| \bm{x} \|}_2 = {\| \bm{x}_i \|}_2 = r_0$ holds for $i=1,...,L$).

Since (\ref{ML_est_ref}) is highly nonconvex, the state-of-the-art gradient-type approach in \cite{SHuang}, GPGD, might be subject to performance deterioration if not properly initialized.

\section{ALGORITHM DEVELOPMENT}

\label{Sect_AD}
\begin{figure*}[!t]
	\centering
	\includegraphics[width=7in]{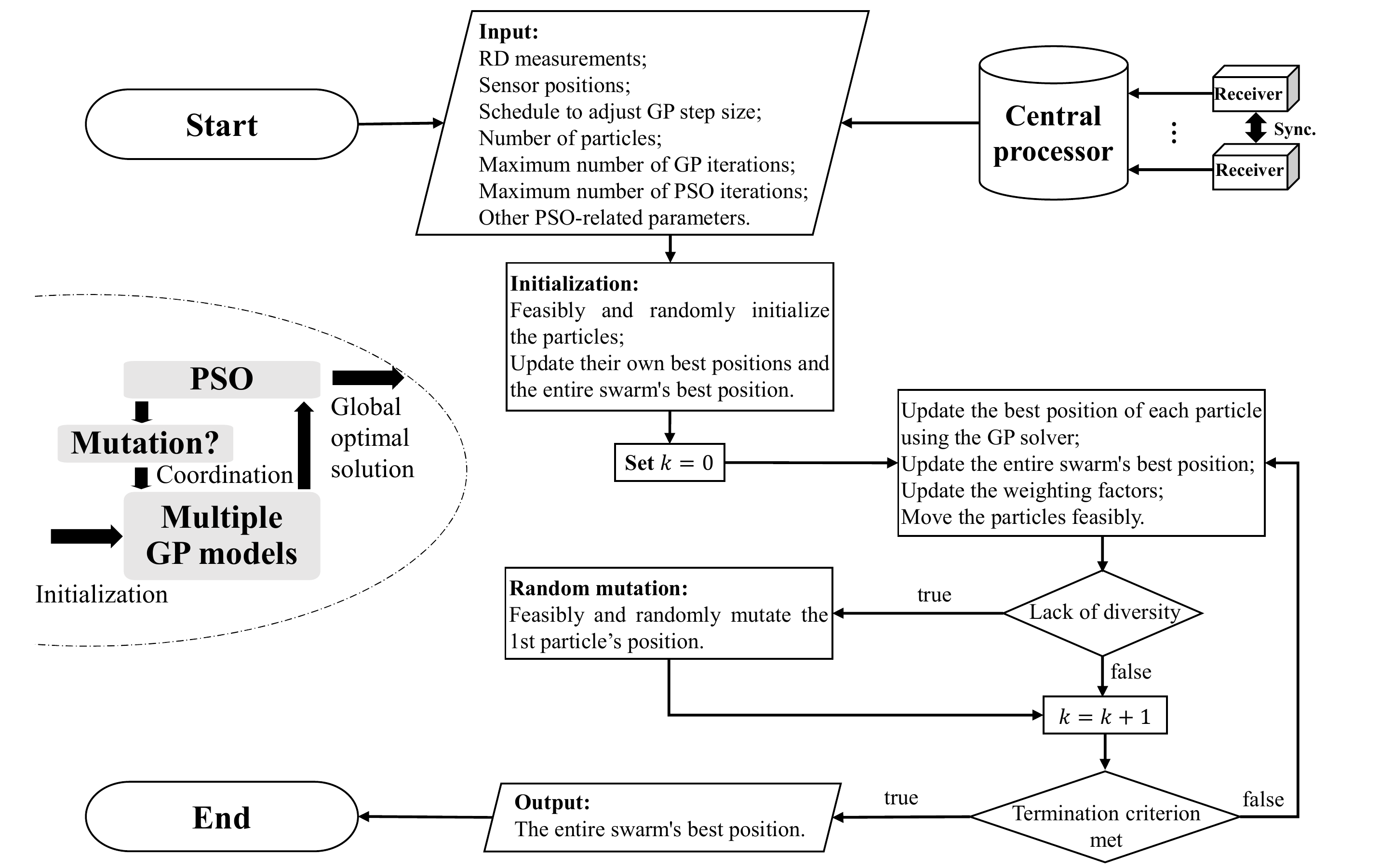}
	\caption{Schematic diagram and flowchart of overall methodology.} 
	\label{FIG_BD}
\end{figure*}

In this section, the globally optimized CGP framework (schematized on the left-hand side of Fig. \ref{FIG_BD}) is presented for coping with (\ref{ML_est_ref}). To put it briefly, there are two components constituting the hierarchy, a PSO metaheuristic at the higher level and multiple critical point finding GP models below. The higher-level collaboration mechanisms contribute to coordinating the deterministic solutions at the lower level and, thereby, guiding the global search process. In the interaction between the two components, mutations are introduced to ensure the diversity of PSO. A more detailed bottom-up explanation is given as follows.

Each of the GP individuals corresponds to an instance of Algorithm \ref{Algorithm1} with its own initial state.

\begin{algorithm}[t]
	\SetAlgoLined
	\caption{Critical Point Finding GP Model.}
	\label{Algorithm1}
	\KwIn{$\bm{r}$, $\{\bm{x}_i|i=1,...,L\}$, $\rho(\cdot)$, and $N_{\max}^{\prime}$.}
	
	{
		
		\textbf{Initialize:} ${\bm{x}}^{(0)} \in \Omega$ randomly, and $\tau^{(0)} = {10}^{-4}$.
		
		\For{$k=0,1,\cdots$}{
		
			$\bm{x}^{(k+1)} \leftarrow \Pi_{\Omega}(\bm{x}^{(k)} - \tau^{(k)} \cdot \nabla_{\bm{x}} g(\bm{x}^{(k)}))$;

			$\tau^{(k+1)} \leftarrow \rho (\tau^{(k)})$;
			
			\textbf{Stop} if $k+1$ reaches a predefined upper limit $N_{\max}^{\prime}$ or $\{ g(\bm{x}^{(k)}) \}$ converges to a reasonable lower level.
		
		}
		
		$\tilde{\bm{x}} \leftarrow \bm{x}^{(k+1)}$;

	}
	\KwOut{Estimate of source location $\tilde{\bm{x}}$.}
	\textbf{Function Handle:} \texttt{GP}$(\{r_{i,1}\}, \{\bm{x}_i\}, \rho(\cdot), N_{\max}^{\prime},{\bm{x}}^{(0)})$
\end{algorithm}

Specifically, we have for the gradient calculations \cite{SHuang}
\begin{align}
	\nabla_{\bm{x}}g = \tfrac{\partial g}{\partial \bm{x}} = -2(\bm{E} \tfrac{\partial \bm{P}^{\prime}}{\partial \bm{\beta}} \tfrac{\partial \bm{\beta}}{\partial \bm{x}})^T \bm{\Sigma}^{-1} (\bm{r} - \bm{E} \bm{P}^{\prime}).
\end{align}
While $\tfrac{\partial \bm{\beta}}{\partial \bm{x}} = [(\tfrac{\partial \beta_1}{\partial \bm{x}})^T,...,(\tfrac{\partial \beta_L}{\partial \bm{x}})^T]^T \in \mathbb{R}^{L \times 3}$ is obtained based on (\ref{2ndeq_cons}) as
\begin{align}
	\tfrac{\partial \beta_i}{\partial \bm{x}} = - \bm{x}_i^T \big/ \big( r_0 \sin \big( \tfrac{D_i}{r_0} \big) \tfrac{d D_i}{d \beta_i} \big),~~i = 1,...,L,
\end{align}
the computation of $\tfrac{d D_i}{d \beta_i}$ (for $i=1,...,L$) and $\tfrac{\partial \bm{P}^{\prime}}{\partial \bm{\beta}}$ as per the basic QP model in (\ref{QPmdl}) are rather straightforward and therefore omitted. Given certain $\bm{x}^{(k)}$, the ray takeoff angles at the $k$th iteration, $\{ \beta_{i}^{(k)} \}$, are produced by solving $2r_0 \sin \big( \tfrac{D_i(\beta_i)}{2r_0} \big) = {\| \bm{x}_i - \bm{x}^{(k)} \|}_2$ for $\beta_1,...,\beta_L$. This can be achieved with ease by simple root-finding algorithms \cite{UMAscher}, making use of the relationship between $D_i$ and $\beta_i$ shown in Section \ref{Sect_PF}.

As for the gradient projections, instead of directly performing the point-to-set mapping $\Pi_{\Omega}(\bm{z})$ by its definition \cite{HAttouch}: $\Pi_{\Omega}(\bm{z}) := \arg \min_{\bm{x} \in \Omega} {\| \bm{x} - \bm{z} \|}_2$, we follow \cite{SHuang} to exploit the structure of the sets $\mathcal{S}$ and $\mathcal{T}$, and employ alternating projections (APs) between them:
\begin{subequations}{\label{onto2sets}}
\begin{align}
	&\bm{y}_{(j)} \leftarrow \Pi_{\mathcal{T}}(\bm{x}_{(j)}) := \arg \min_{\bm{y} \in \mathcal{T}} {\| \bm{y} - \bm{x}_{(j)} \|}_2,\label{ontoP}\\
	&\bm{x}_{(j+1)} \leftarrow \Pi_{\mathcal{S}}(\bm{y}_{(j)}) := r_0 \cdot \bm{y}_{(j)} / {\| \bm{y}_{(j)} \|}_2,\label{ontoS}
\end{align}
\end{subequations}
where the superscript is dropped for the moment for notational simplicity. Dealing with the subproblem (\ref{ontoP}) is equivalent to tackling the following quadratic program:
\begin{equation}{\label{QP_prob}}
	\min_{\bm{y}}~\tfrac{1}{2}{\| \bm{y} - \bm{x}_{(j)} \|}_2^2,~~\textup{s.t.}~\bm{A} \bm{y} \leqq \bm{b},
\end{equation}
where $\bm{A} = [-\bm{x}_1,...,-\bm{x}_L,\bm{x}_1,...,\bm{x}_L]^T \in \mathbb{R}^{2L \times 3}$, $\bm{b} \in \mathbb{R}^{2L}$, $[\bm{b}]_{i} = -r_0^2 + 2 r_0^2 \sin^2 \big( \tfrac{D_i(0)}{2r_0} \big)$ (for $i = 1,...,L$), $[\bm{b}]_{i+L} = r_0^2 - 2 r_0^2 \sin^2 \big( \tfrac{D_i(\beta^{\textup{U}})}{2r_0} \big)$ (for $i = 1,...,L$). There do exist many efficient solvers for quadratic programming, as it is a classical problem on which the research may date back to the 1950s (e.g., the Frank-Wolfe algorithm \cite{MFrank} and Hildreth's method \cite{CHildreth}).

Note that the local linear convergence of AP with nonconvexity to a point in the intersection of two sets is guaranteed under very mild regularity conditions \cite{ASLewis}. Moreover, in our simulations, it is observed that a single run of AP already yields satisfying results.

\textit{Remark on Convergence Properties of GP}\textit{:} Analogous to \cite{SHuang}, the general analytical results for GP methods in \cite{HAttouch} can be leveraged to study the convergence of Algorithm \ref{Algorithm1}. Let $\hat{\Omega}$ be a (nonempty) closed and bounded subset of $\Omega$. It is not hard to verify that $g$ is a differentiable function whose gradient is $\zeta$-Lipschitz continuous on $\hat{\Omega}$, which, in other words, means there exists a positive real constant $\zeta$, such that ${\| \nabla_{\bm{x}}g(\bm{x}) - \nabla_{\bm{y}}g(\bm{y}) \|}_2 \leq \zeta {\| \bm{x} - \bm{y} \|}_2$ holds for all real-valued $\bm{x}, \bm{y} \in \hat{\Omega}$. Based on \cite[Theorems 5.1 and 5.3]{HAttouch}, for a given $\epsilon \in (0, 1/(2 \zeta))$ and a sequence of step sizes $\{ \tau^{(k)} \}$ that satisfy $\epsilon < \tau^{(k)} < (1 / \zeta) - \epsilon$, the sequence of $\{ \bm{x}^{(k)} \in \hat{\Omega} \}$ generated by GP in Algorithm \ref{Algorithm1} converges to a critical point of (\ref{ML_est_ref}).

\begin{algorithm}[t]
	\SetAlgoLined
	\caption{Globally Optimized CGP Method.}
	\label{Algorithm2}
	\KwIn{$\bm{r}$, $\{\bm{x}_i|i=1,...,L\}$, $\rho(\cdot)$, $N_{\textup{Ptcl}}$, $N_{\max}^{\prime}$, $N_{\max}$, $w_{\max}$, $w_{\min}$, $c_1$, $c_2$, and $\gamma_{\textup{Div}}$.}
	
	{
		
		\textbf{Initialize:} 
		
		1) ${\bm{p}}_{j}^{(0)} \in \Omega$ randomly for $j = 1,...,N_{\textup{Ptcl}}$; 
		
		2) $\bm{p}_{j}^{\textup{b}} \leftarrow {\bm{p}}_{j}^{(0)}$ for $j = 1,...,N_{\textup{Ptcl}}$; 
		
		3) $i \leftarrow \arg \min_{i \in \{1,...,N_{\textup{Ptcl}}\}} g(\bm{p}_{i}^{\textup{b}})$; $\bm{p}^{\textup{g}} \leftarrow \bm{p}_{i}^{\textup{b}}$.
		
		\For{$k=0,1,\cdots$}{

			\For{$j=1,\cdots,N_{\textup{Ptcl}}$}{

			$\tilde{\bm{p}}_{j} \leftarrow$ \texttt{GP}$(\{r_{i,1}\}, \{\bm{x}_i\}, \rho(\cdot), N_{\max}^{\prime},{\bm{p}}_{j}^{(k)})$;

			\If{$g(\tilde{\bm{p}}_{j}) < g(\bm{p}_{j}^{\textup{b}})$}{

			$\bm{p}_{j}^{\textup{b}} \leftarrow \tilde{\bm{p}}_{j}$;

			}

			}

			$i \leftarrow \arg \min_{i \in \{1,...,N_{\textup{Ptcl}}\}} g(\bm{p}_{i}^{\textup{b}})$; 
			
			$\bm{p}^{\textup{g}} \leftarrow \bm{p}_{i}^{\textup{b}}$;

			$w^{(k)} \leftarrow w_{\max} - ( \tfrac{w_{\max} - w_{\min}}{N_{\max}} ) \cdot k$;

			\For{$j=1,\cdots,N_{\textup{Ptcl}}$}{

			$\bm{v}_{j}^{(k+1)} \leftarrow w^{(k)} \cdot \bm{v}_{j}^{(k)} + c_1 \cdot \texttt{rand} \cdot (\bm{p}_{j}^{\textup{b}} - \bm{p}_{j}^{(k)}) + c_2 \cdot \texttt{rand} \cdot (\bm{p}^{\textup{g}} - \bm{p}_{j}^{(k)})$;

			$\bm{p}_{j}^{(k+1)} \leftarrow \Pi_{\mathcal{S}}(\Pi_{\mathcal{T}}(\bm{p}_{j}^{(k)} + \bm{v}_{j}^{(k+1)}))$;

			}

			\If{$\textup{Div}_{N_{\textup{Ptcl}}}(\bm{p}^{\textup{g}}) < \gamma_{\textup{Div}}$}{

			$\bm{p}_{1}^{(k+1)} \leftarrow \mathcal{M}(\bm{p}_{1}^{(k+1)}) \in \Omega$;

			}

			\textbf{Stop} if $k+1$ reaches a predefined upper limit $N_{\max}$ or $\{ \bm{p}^{\textup{g}} \}$ converges to a reasonable lower level.

	    }
		
		$\tilde{\bm{x}} \leftarrow \bm{p}^{\textup{g}}$;

	}
	\KwOut{Estimate of source location $\tilde{\bm{x}}$.}
\end{algorithm}

The PSO is a population-based computational method that searches for the global optimum of a function by mimicking the social behaviors of bird flocking and fish schooling \cite{FJavidrad}. The population of candidate solutions, a.k.a. the swarm of particles, is initially randomly distributed across the search space, and then moved according to simple updating rules for each particle's position and velocity. With the equations of movement depending on both their own best positions in the region and the best position that the entire swarm ever reached, the particles are guided towards the global optimum.

Let us assume the search space is in $\mathbb{R}^{N}$, and denote the position of the $j$th particle in the $k$th iteration, the best position of the $j$th particle in the region, and the entire swarm's best position so far by $\bm{p}_{j}^{(k)} = [p_{j,1}^{(k)},...,p_{j,N}^{(k)}]^T \in \mathbb{R}^{N}$, $\bm{p}_{j}^{\textup{b}} = [p_{j,1}^{\textup{b}},...,p_{j,N}^{\textup{b}}]^T \in \mathbb{R}^{N}$, and $\bm{p}^{\textup{g}} = [p_{1}^{\textup{g}},...,p_{N}^{\textup{g}}]^T \in \mathbb{R}^{N}$, respectively. Note that the superscript hereinafter denotes the PSO iteration number (rather than the GP's as in Algorithm \ref{Algorithm1}). The updating rules of a frequently used PSO algorithm are given as \cite{FJavidrad}:
\begin{subequations}{\label{PSO_v}}
\begin{align}
	&\bm{v}_{j}^{(k+1)} \leftarrow w^{(k)} \cdot \bm{v}_{j}^{(k)} + c_1 \cdot \texttt{rand} \cdot (\bm{p}_{j}^{\textup{b}} - \bm{p}_{j}^{(k)}) \nonumber\\
	&~~+ c_2 \cdot \texttt{rand} \cdot (\bm{p}^{\textup{g}} - \bm{p}_{j}^{(k)}),\label{PSO_va}\\
	&\bm{p}_{j}^{(k+1)} \leftarrow \bm{p}_{j}^{(k)} + \bm{v}_{j}^{(k+1)},\label{PSO_vb}
\end{align}
\end{subequations}
where $w^{(k)} = w_{\max} - ( \tfrac{w_{\max} - w_{\min}}{N_{\max}} ) \cdot k$ is the weighting factor introduced for a better control of the search scope, $w_{\max}$ and $w_{\min}$ are two predefined limits, and $c_1, c_2 \in \mathbb{R}$ are the cognitive and social parameters, respectively, adjusting the particle's steps taken in a movement. In our simulations, we follow \cite{FJavidrad} to adopt the near optimum setting of parameters by letting $w_{\max} = 0.6$, $w_{\min} = 0.15$, $c_1 = 1.8$, and $c_2 = 1$.

Nevertheless, despite the straightforward usability of (\ref{PSO_v}), such a simple procedure might suffer from the premature convergence (especially if in the presence of multimodality \cite{SCEsquivel}). There is thus the need for improvement in the swarm diversity, which can be achieved by introducing a mutation operator in the implementation of PSO \cite{PSAndrews}: $\bm{p}_{j}^{\textup{n}} \leftarrow \mathcal{M}(\bm{p}_{j})$, where $\bm{p}_{j}^{\textup{n}} \in \mathbb{R}^{N}$ means the mutation-generated new position on the basis of the current position $\bm{p}_{j} \in \mathbb{R}^{N}$. The step is invoked next to (\ref{PSO_vb}) when the diversity measure: $\textup{Div}_{N_{\textup{Ptcl}}}(\bm{p}^{\textup{g}}) = \tfrac{1}{N_{\textup{Ptcl}}} \sum_{j = 1}^{N_{\textup{Ptcl}}} {\big\| \bm{p}_{j}^{(k+1)} - \bm{p}^{\textup{g}} \big\|}_2$ is less than a predefined tolerance $\gamma_{\textup{Div}}$, which implies a dense swarm of particles. Here we adopt the random mutation operator \cite{PSAndrews} for $\mathcal{M}(\cdot)$.

Another point to consider is that the PSO cannot directly handle the constraints in the formulation (requiring either the penalty method or a feasibility-based rule \cite{FJavidrad}). To overcome the defect and further reduce the possibility of convergence to undesirable local optima, we propose a hybrid stochastic and deterministic scheme with the responsibilities for local search being handed over to multiple lower-level GP models. In so doing, higher efficiency can be attained in the sense that searches for the global optimum in CGP are carried out among only a finite number of critical point solutions (instead of an infinite number of candidates within the search space in the basic PSO). Concretely, updating the initial states of $N_{\textup{Ptcl}}$ GP models via the PSO (with $N = 3$) establishes our CGP framework in Algorithm \ref{Algorithm2}, where the AP is applied once in updating each particle's position so that it falls into the problem domain $\Omega$. It is worth noting that a random mutation operator is employed to regenerate the position of the first particle in case of lack of swarm diversity.

A flowchart is depicted in Fig. \ref{FIG_BD} to better visualize the overall methodology.

\textit{Remark on Convergence Properties of CGP}\textit{:} The general convergence proofs for random search techniques in \cite{FJSolis} can be extended to our case. There are two conditions sufficient to ensure the convergence of a conceptual global search algorithm, which begins with a point ${\bm{p}}^{(0)} \in \Omega$ (a subset of $\mathbb{R}^{N}$), and seeks a point therein minimizing $g: \mathbb{R}^{N} \rightarrow \mathbb{R}$ on $\Omega$, by iteratively generating the random sample $\bm{\xi}^{(k)}$ from the sample space $(\mathbb{R}^{N},\mathcal{B},\mu^{(k)})$, setting $\bm{p}^{(k+1)} \leftarrow D(\bm{p}^{(k)},\bm{\xi}^{(k)})$, and choosing $\mu^{(k)}$. Here, $\mu^{(k)}$ denotes the (conditional) probability measure corresponding to certain distribution function defined on $\mathbb{R}^{N}$, and the map $D$ is with domain $\Omega \times \mathbb{R}^N$ \cite{FJSolis}. The first condition to be verified is: $g(D(\bm{p},\bm{\xi})) \leq g(\bm{p})$, and $g(D(\bm{p},\bm{\xi})) \leq g(\bm{\xi})$ if $\bm{\xi} \in \Omega$. This is obviously satisfied by our CGP framework, as finding the swarm's best position ever experienced in PSO, by its nature, monotonically non-increases the below-bounded objective function $g$, and the same goes for every deterministic GP model initialized to a PSO-generated random state. The second condition is stated as $\prod_{k=0}^{\infty} (1 - \mu^{(k)}(Y)) = 0$ holds for any Borel subset $Y$ of $\Omega$ with $v(Y) > 0$. In other words, it means the zero possibility of repeatedly missing any given subset $Y$ of $\Omega$ with positive volume when generating $\bm{\xi}^{(k)}$ \cite{FJSolis}. Because a random mutation operator has been introduced to re-initialize the first GP model in $\Omega$, it follows that $v(\Omega \cap M^{(k)}) = v(\Omega)$ with $M^{(k)}$ being the support of $\mu^{(k)}$ (the smallest closed subset of $\mathbb{R}^{N}$ with $\mu^{(k)} = 1$), namely, the solution space is covered by the search space. Therefore, the second condition is satisfied as well. According to \cite[Convergence Theorem]{FJSolis}, if we suppose that $g$ and $\Omega$ are both measurable, it is then foreseeable that CGP will produce a solution falling into a small volume of search space surrounding the global optimum, as long as sufficient quantities of GP models and PSO iterations are guaranteed. In fact, our simulation results suggest that two GP individuals and five PSO iterations are already enough for CGP to converge to the global optimum in most scenarios.

From Algorithm \ref{Algorithm2} and Fig. \ref{FIG_BD}, it is not difficult to see that the computational cost of the CGP method is dominated by those involved in invoking the lower-level GP models, where the deterministic local optimizers lie. The most computationally extensive procedure in the GP algorithm, on the other hand, is the AP in (\ref{onto2sets}) between two sets (more precisely, the quadratic programming (\ref{QP_prob}) that computes the projection onto $\mathcal{T}$). As we consider using approximation schemes to deal with the corresponding quadratic program and performing the AP only once at each GP iteration, the implementation of a GP model leads to $\mathcal{O}(N_{\textup{GP}} N_{\textup{Quad}})$ complexity, where $N_{\textup{GP}}$ and $N_{\textup{Quad}}$ denote the numbers of iterative steps needed by the GP algorithm and the quadratic programming solver, respectively.

Since the contribution of our work lies mainly in the novel re-initialization scheme based on the PSO, we are likewise interested in the specific improvements over the typical initialization tactics in the literature. Utilizing the coordinate descent algorithm and the Newton-like method, the authors of \cite{TWang} have designed a so-called ``warm-start'' procedure \cite{SHuang} to select the initial point for the location estimate in their three-step heuristic. However, as discussed in \cite{SHuang}, ray takeoff angles estimated by the penalized formulation in \cite{TWang} may exceed the limit angle associated with the skip distance. This will in turn result in an inappropriate starting point for the subsequent estimation of source location and therefore the degraded positioning performance. Another ready-made initialization scheme is the randomized one adopted by the GPGD method in \cite{SHuang}. In a similar fashion to ours in the individual GP model, it simply randomly picks a point from the problem domain $\Omega$ at the very beginning of algorithm implementation. Although the ambiguity associated with the ray takeoff angles has been resolved by means of constraints akin to (\ref{PolytopeT}), the random initialization in \cite{SHuang} still has a major drawback: it does not assure the gradient-type GPGD approach a close-enough starting point. As demonstrated by our simulation results in Section \ref{Sect_NR}, the initialization challenge remains for \cite{SHuang} and the positioning accuracy of GPGD is generally rather poor. In contrast, such issues are well tackled by the re-initialization scheme presented herein, in the sense that the unrealistic ray takeoff angles are ruled out via (\ref{PolytopeT}), and the shortcoming of GPGD is remedied using the metaheuristic of PSO.

\section{SIMULATIONS AND PRACTICAL ISSUES}

\label{Sect_NR}
In this section, we conduct numerical experiments to evaluate the performance of CGP in comparison with the state-of-the-art GPGD method in \cite{SHuang}, and discuss the challenges when it is applied for TDOA HF source localization in practice. All the simulations are carried out on a laptop with 16 GB memory and a 4.7 GHz CPU.

\begin{table}[!t]
	\renewcommand{\arraystretch}{1}
	\caption{Deployment of HF Source and Sensors}
	\label{table_QPTDOA}
	\centering
	\begin{tabular}{|c|c|c|c|}
		\hline
		\bfseries Place & \bfseries Description & \bfseries Lat. \& Long. & \bfseries Takeoff Angle\\
		\hline
		Freiburg & Source & ${48.00}^{\circ}$, ${7.84}^{\circ}$ & N/A\\
		\hline
		Berlin & Sensor 1 & ${52.52}^{\circ}$, ${13.41}^{\circ}$ & ${33.77}^{\circ}$\\
		\hline
		Paris & Sensor 2 & ${48.86}^{\circ}$, ${2.35}^{\circ}$ & ${57.14}^{\circ}$\\
		\hline
		Cambridge & Sensor 3 & ${52.20}^{\circ}$, ${0.12}^{\circ}$ & ${29.09}^{\circ}$\\
		\hline
		Vienna & Sensor 4 & ${48.21}^{\circ}$, ${16.37}^{\circ}$ & ${34.17}^{\circ}$\\
		\hline
		Amsterdam & Sensor 5 & ${52.37}^{\circ}$, ${4.90}^{\circ}$ & ${42.57}^{\circ}$\\
		\hline
	\end{tabular}
\end{table}

\begin{figure}[!t]
	\centering
	\includegraphics[width=3.2in]{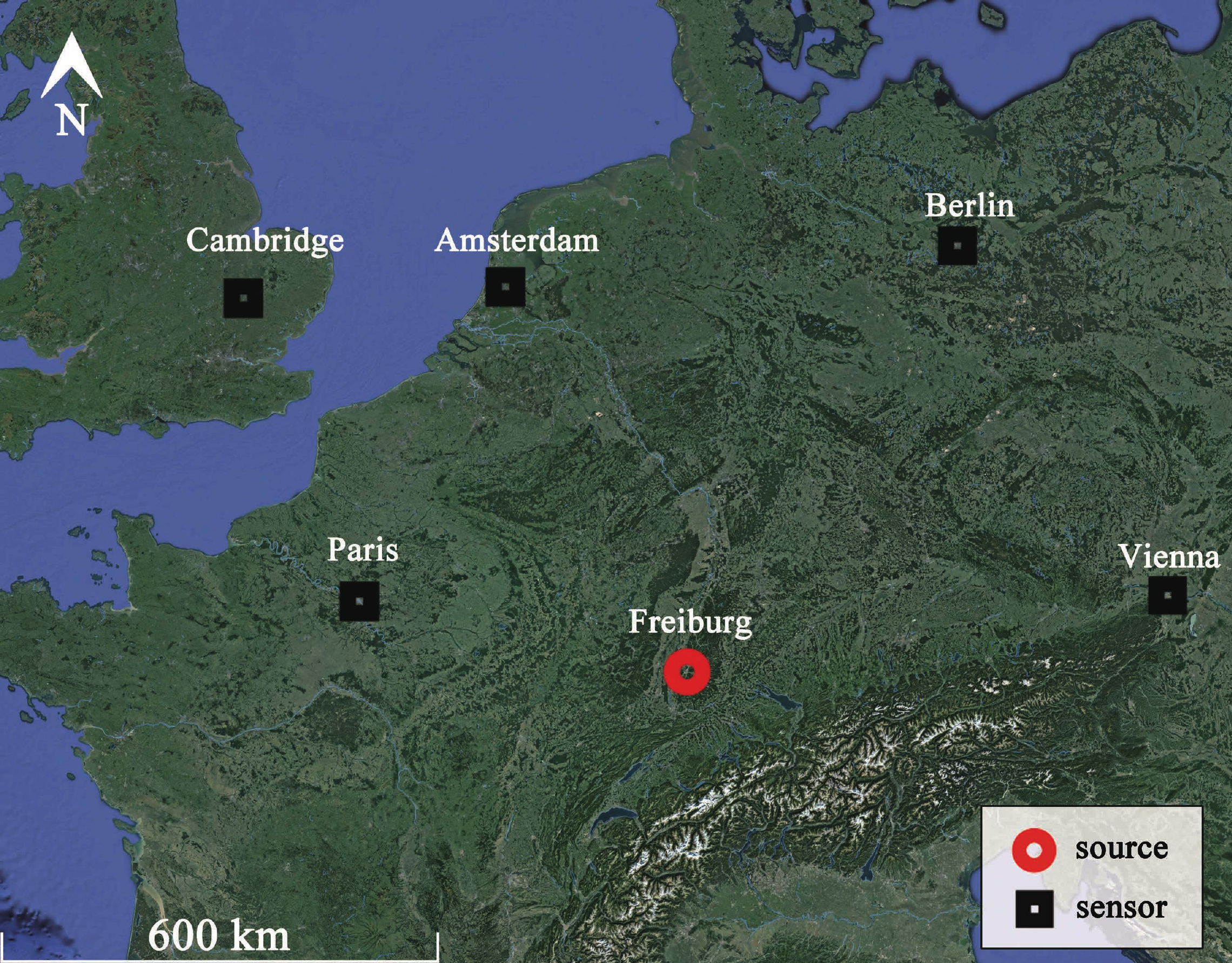}
	\caption{Illustration of involved geographical region in simulations.} 
	\label{geomap}
\end{figure}

We consider an over-the-horizon radar (OTHR) system with a single HF source and $L=5$ sensors being deterministically deployed on the ground. The geodetic coordinates (latitude and longitude) are found on Google Earth \cite{GoogleEarth} and listed in Table \ref{table_QPTDOA}, and we convert them into the geocentric Cartesian coordinates for mathematical calculations. The involved geographical region is illustrated in Fig. \ref{geomap}. OTHR-related parameters are fixed as $r_m = 6650$ km, $r_b = 6550$ km, $y_m = 100$ km, $f = 11$ MHz, and $f_c = 10$ MHz. As a consequence, we have $\beta^{\textup{U}} = {60.43}^{\circ}$ in such a configuration. The corresponding ray takeoff angles are also provided in Table \ref{table_QPTDOA}. Regarding the parameters associated with the GP models, $\rho(\cdot)$ is kept the same as that in \cite{SHuang}, and we set $N_{\max}^{\prime}$ to $10^{4}$. Unless otherwise indicated, the whole set of 5 sensors are employed for localization and the tolerance for the swarm diversity is fixed at $200$ km. Furthermore, we follow the general setting of \cite{SHuang} to randomly and feasibly initialize the GPGD method in the simulations.

\begin{figure}[!t]
	\centering
	\includegraphics[width=3.2in]{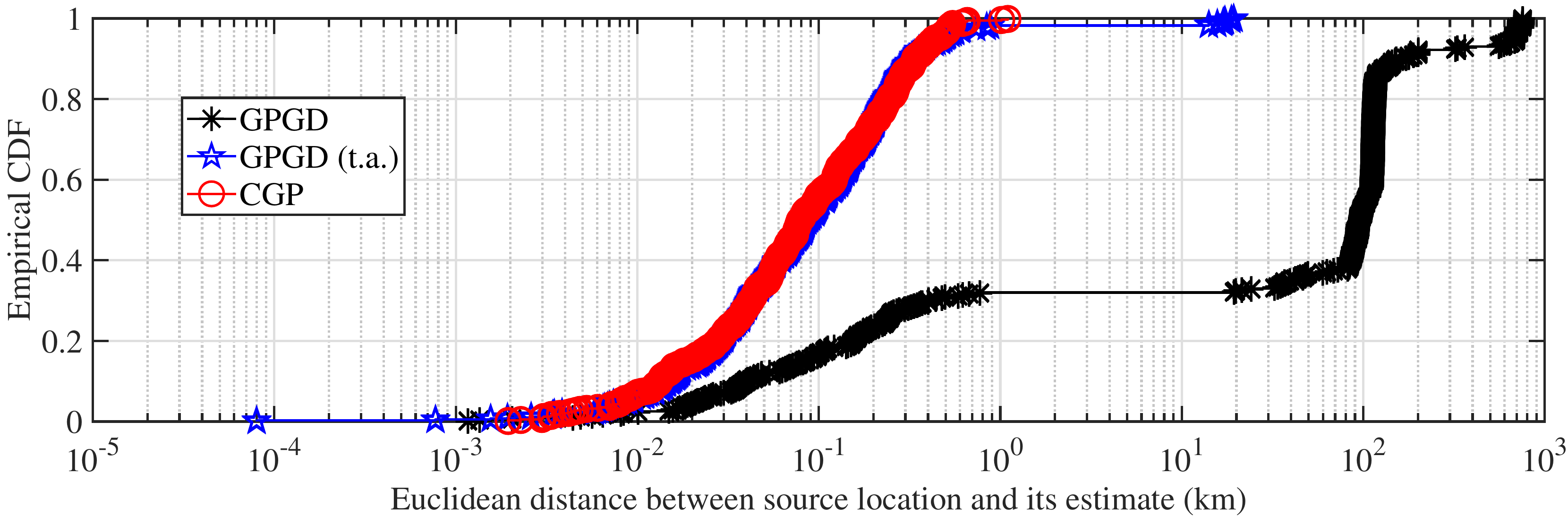}
	\caption{Empirical CDF of positioning error.} 
	\label{CDFPlot}
\end{figure}

As a simple but representative example, we start with assessing the localization accuracy of the proposed CGP scheme in the scenario where the Gaussian disturbances $\{n_i | i = 1,...,L\}$ are assumed to be independent and identically distributed (i.i.d.) with constant standard deviation $\sigma$. Based on 100 Monte Carlo (MC) instances generated for each noise level, Fig. \ref{CDFPlot} shows the empirical cumulative distribution functions (CDFs) of the positioning error ${\left\|\tilde{\bm{x}} - \bm{x}\right\|}_2$ for GPGD and CGP when $\sigma \in \{ 10, 40, 70, 100 \}$ m, $N_{\textup{Ptcl}} = 2$, and $N_{\max} = 5$. We see that the randomly initialized GPGD method in general does not exhibit reliable positioning performance, i.e., more than 60\% of the samples gives rise to errors larger than 1 km (even several hundreds of kilometers in the worst case). Additionally, we follow \cite{SHuangCode} (source code for \cite{SHuang} shared by its authors) to include ${\left\|\tilde{\bm{x}} - \bm{x}\right\|}_2 \leq 20$ km into the experimental protocol and accordingly the new outcome of GPGD in Fig. \ref{CDFPlot}, where ``threshold applied'' is abbreviated to ``t.a.''. As one may see, the criterion does not change the situation that there still exist seriously erroneous estimates below the threshold (with positioning errors ranging from 1 km to 20 km), not to mention its practical unavailability. On the contrary, our CGP algorithm can persistently deliver location estimates satisfying ${\left\|\tilde{\bm{x}} - \bm{x}\right\|}_2 \leq \delta$ without presetting any similar threshold, no matter where the starting point is chosen in the problem domain $\Omega$. Here, $\delta$ amounts to a value around 1 km.

\begin{table}[!t]
	\renewcommand{\arraystretch}{1}
	\caption{Performance Comparison of Estimators in i.i.d. Noise}
	\label{table_RMSE}
	\centering
	\begin{tabular}{|c|c|c|c|c|}
		\hline
		\bfseries Estimator & GPGD & GPGD (t.a.) & CGP & CRLB \\
		\hline
		\bfseries RMSE \mdseries (m) & 205259.5 & 2269.7 & 206.1 & 66.0 \\
		\hline
		\bfseries RGE & 28.46\% & 0.31\% & 0.31\textperthousand & 0.09\textperthousand \\
		\hline
	\end{tabular}
\end{table}

Table \ref{table_RMSE} then compares the localization performance of GPGD, GPGD (t.a.), and CGP in this simulation in terms of the root-mean-square error (RMSE) and the RGE \cite{AJain3}, defined as $\textup{RMSE} = \sqrt{\tfrac{1}{N_{\textup{MC}}}\sum_{i=1}^{N_{\textup{MC}}}{{\left\|\tilde{\bm{x}}^{\{i\}} - \bm{x}\right\|}_2^2}}$ and $\textup{RGE} = \textup{RMSE}/\max_{i \in \{ 1,...,L \}} D_i$, respectively, where $\tilde{\bm{x}}^{\{i\}} \in \mathbb{R}^{3}$ is the source location estimate associated with the $i$th MC sample, and $N_{\textup{MC}}$ the total of MC trials. The Cram\'{e}r-Rao lower bound (CRLB) for $\bm{x}$, originally derived by the authors of \cite{TWang}, is also included in Table \ref{table_RMSE} to offer an accuracy benchmark. It is observed that CGP produces the lowest RMSE and RGE values among the estimators GPGD, GPGD (t.a.), and CGP, corresponding to the closest one to the theoretical limit of CRLB.

\begin{figure}[!t]
	\centering
	\includegraphics[width=3.2in]{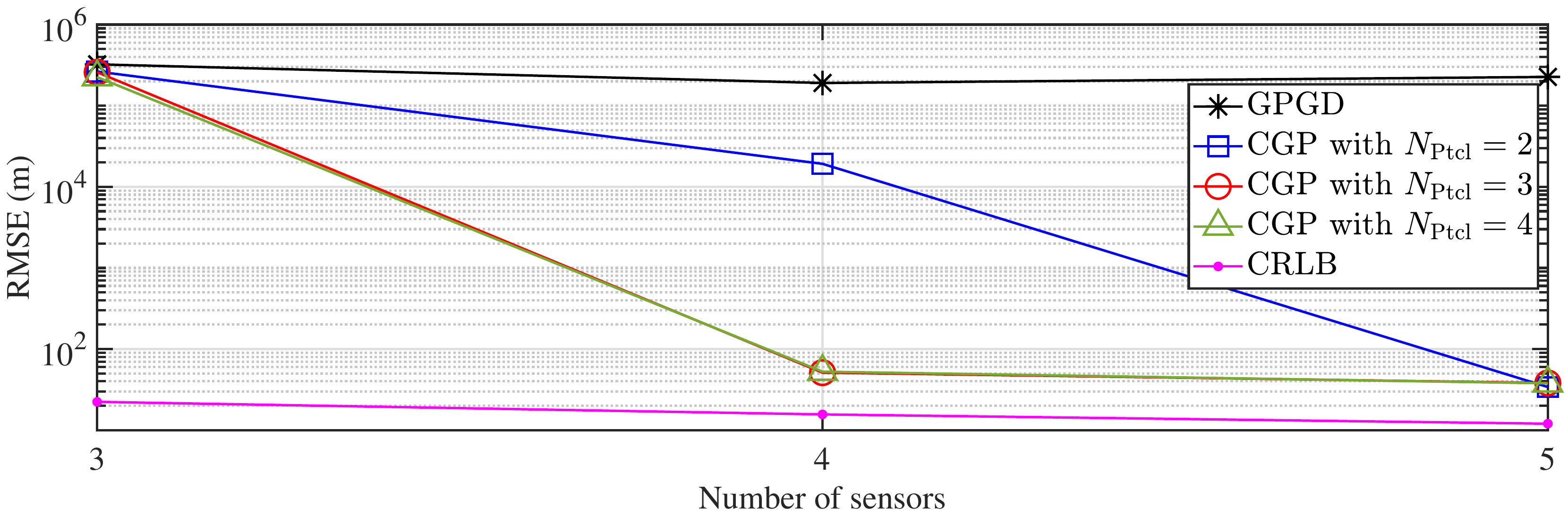}
	\caption{RMSE versus number of sensors.} 
	\label{RMSEvsSensorNum}
\end{figure}

Next, the minimum sensor number requirement for the proposed CGP algorithm is numerically analyzed. Assuming that $L$ sensors starting from Sensor 1 in Table \ref{table_QPTDOA} are used for geolocation (the same indexing applies to all the scenarios hereinafter with $L < 5$), Fig. \ref{RMSEvsSensorNum} plots the RMSE versus $L \in \{ 3,4,5 \}$ when $\sigma = 10$ m, $N_{\textup{Ptcl}} \in \{ 2,3,4 \}$, and $N_{\max} = 5$. In each case of $L \in \{ 3,4,5 \}$, 100 MC trials are performed. We see 4 sensors will be sufficient for CGP to locate the source, provided that the number of particles (namely, GP models) is larger than or equal to 3. Otherwise, at least 5 sensors would be recommended, which ensures the localizability of CGP even with two particles. The GPGD, in comparison, fails to locate the source under the same circumstances.

\begin{figure}[!t]
	\centering
	\includegraphics[width=3.2in]{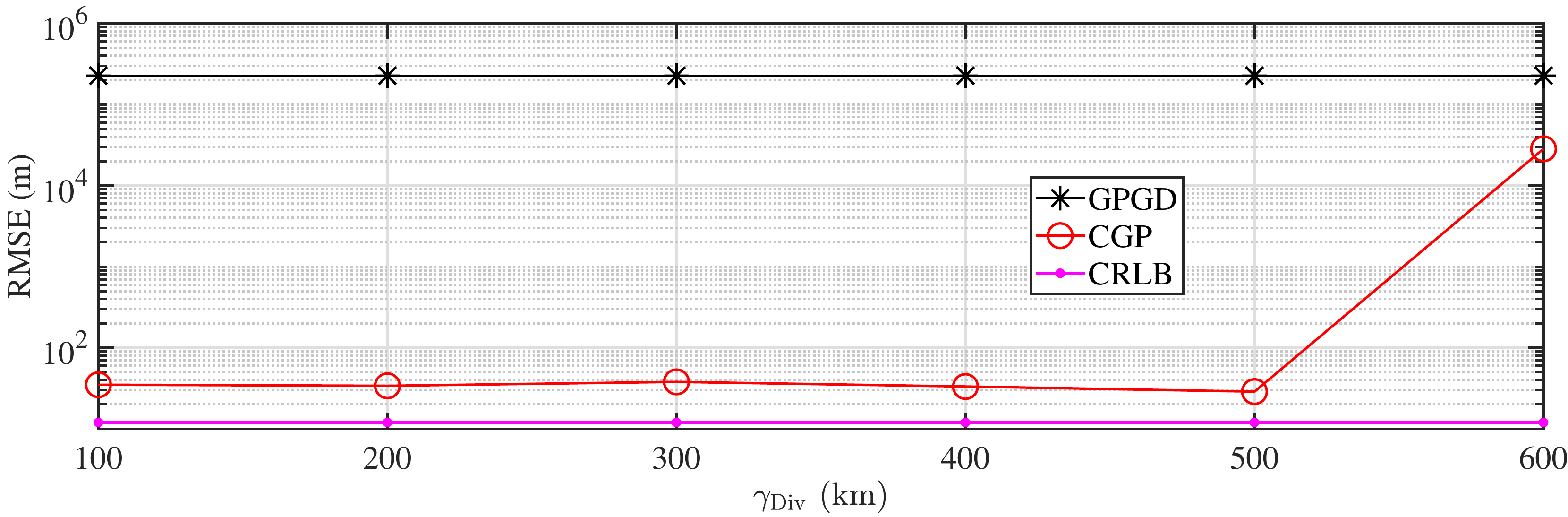}
	\caption{RMSE versus predefined tolerance for swarm diversity.} 
	\label{RMSEvsDivThre}
\end{figure}

\begin{figure}[!t]
	\centering
	\includegraphics[width=3.2in]{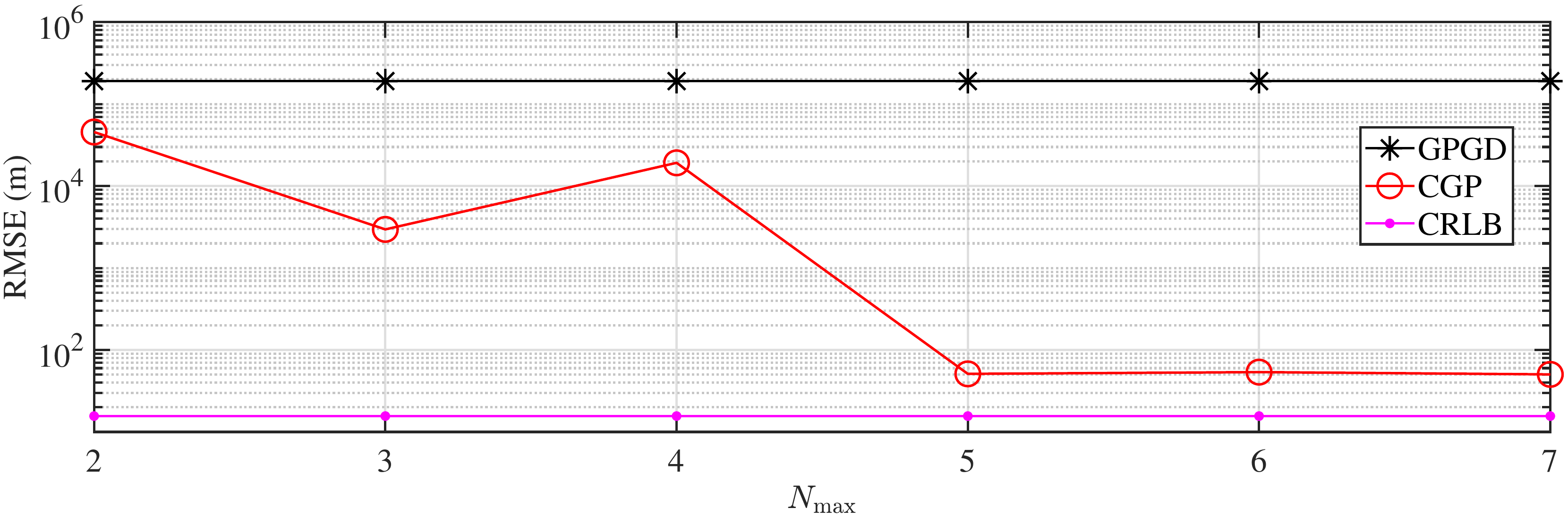}
	\caption{RMSE versus maximum number of PSO iterations.} 
	\label{RMSEvsNmax}
\end{figure}

\begin{figure}[!t]
	\centering
	\includegraphics[width=3.2in]{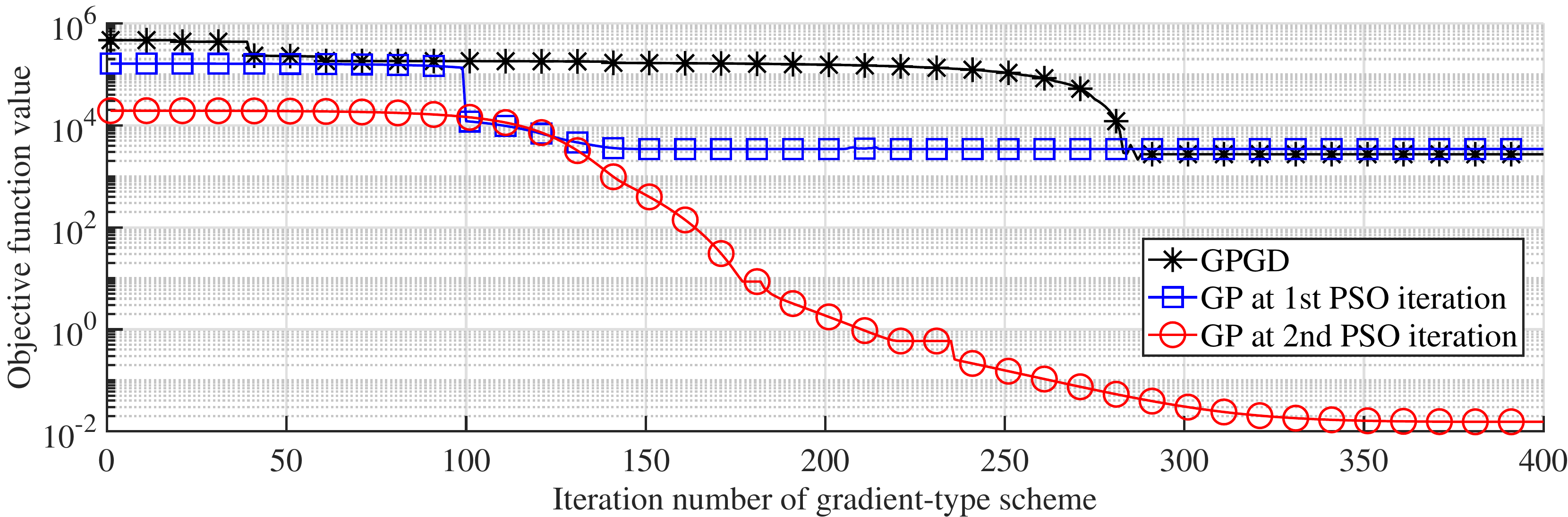}
	\caption{Objective function value versus number of descent iterations in a single MC run when $L = 5$, $\sigma = 10$ m, and $N_{\textup{Ptcl}} = 2$.} 
	\label{ObjPlot}
\end{figure}

While the effect of the number of particles on CGP's performance has been demonstrated in Fig. \ref{RMSEvsSensorNum}, those of the other PSO-related important optimization parameters are studied as follows. By implementing 100 MC runs for each $\gamma_{\textup{Div}}$-realization, Fig. \ref{RMSEvsDivThre} plots the RMSE versus the predefined tolerance for the swarm diversity when $L = 5$, $\sigma = 10$ m, $N_{\textup{Ptcl}} = 2$, and $N_{\max} = 5$. It is seen that the tolerance should not be chosen too large in order to avoid the possible performance deterioration. A threshold of $\gamma_{\textup{Div}} = 500$ km is observed from Fig. \ref{RMSEvsDivThre}. Based on 100 MC trials for each $N_{\max}$-realization, Fig. \ref{RMSEvsNmax} shows the RMSE versus the predefined maximum number of PSO iterations when $L = 4$, $\sigma = 10$ m, and $N_{\textup{Ptcl}} = 3$. As we see, $N_{\max} \geq 5$ is needed for guaranteeing the global optimality of CGP in all MC trials. Nonetheless, it is worth pointing out that even $N_{\max} = 2$ can sometimes be sufficient for CGP to work out the globally optimal solution, e.g., as in the MC instance depicted by Fig. \ref{ObjPlot}. Once again, no meaningful results are yielded by GPGD.

\begin{figure}[!t]
	\centering
	\includegraphics[width=3.2in]{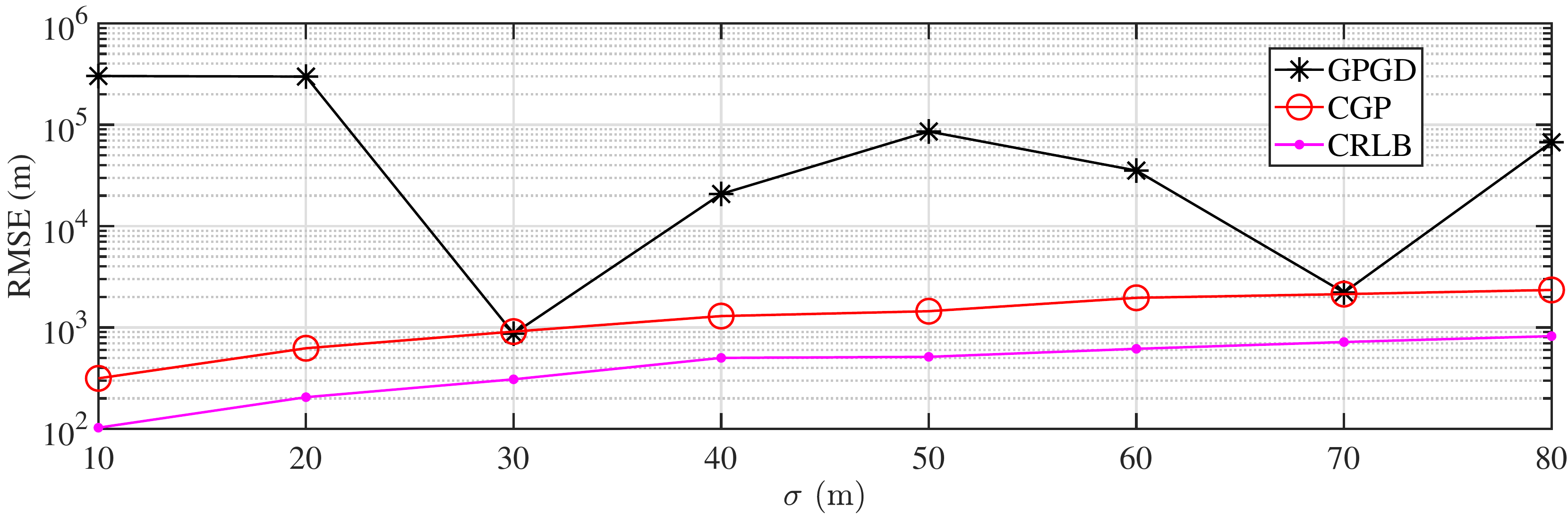}
	\caption{RMSE versus $\sigma$.} 
	\label{RMSEvssigma}
\end{figure}

\begin{table}[!t]
	\renewcommand{\arraystretch}{1}
	\caption{Performance Comparison of Estimators in Noise on Different Scales}
	\label{table_RMSE2}
	\centering
	\begin{tabular}{|c|c|c|c|}
		\hline
		\bfseries Estimator & GPGD & CGP & CRLB \\
		\hline
		\bfseries RMSE \mdseries (m) & 155890.0 & 1541.8 & 528.7 \\
		\hline
		\bfseries RGE & 21.61\% & 0.21\% & 0.07\% \\
		\hline
	\end{tabular}
\end{table}

We proceed now to investigate how the proposed methodology behaves in the more general cases where the measurement noise is no longer identically distributed across all sensors. Based on 100 MC runs for each noise level, Fig. \ref{RMSEvssigma} plots the RMSE versus $\sigma \in [10,80]$ m when $\sigma_{1} = \sigma_{2} = \sigma_{3} = 10 \sigma$, $\sigma_{4} = \sigma_{5} = \sigma$, $N_{\textup{Ptcl}} = 2$, and $N_{\max} = 5$. Similar to Table \ref{table_RMSE}, Table \ref{table_RMSE2} lists the RMSE and RGE of all estimators averaged over $N_{\textup{MC}} = 800$ samples. Note that we do not show the results of GPGD (t.a.) because it is impracticable and does not provide more information here. Despite the gap between the RMSE of CGP and the CRLB, the CGP algorithm overcomes the principal difficulty of initialization and is therefore capable of stably delivering the source location estimates with relatively small errors in all cases. By contrast, the RMSE of GPGD experiences wide and frequent fluctuations as $\sigma$ changes.

To implement the proposed methodology in an operational context, the following practical issues and limitations should be taken into consideration in addition to solving the pure optimization problem (\ref{ML_est_ref}).

\textit{1) Biased Sensor Observations:} As the contamination is common for HF radio signals propagating through the complex ionospheric channel, the sensor-collected TDOA data may contain bias errors apart from the lower-level Gaussian noise in the basic model (\ref{TDOAmdl}). Passing these erroneous data to the algorithm without further treatment can severely degrade the positioning performance. Typically, the negative effects of biased sensor observations are mitigated either when measuring the TDOAs or in the subsequent step of location estimation. For example, carefully filtering the captured signals \cite{AJain2} and choosing the peak with the strongest signal level in cross-correlation \cite{AJain4} are both popularly used schemes that belong to the first category of approaches. The second category, on the other hand, might permit more flexibility. This is because there have been plentiful error-mitigated positioning solutions for the traditional TDOA model in the literature \cite{WXiong3}, extensible to the HF TDOA localization scenarios here upon proper modifications. For instance, one may employ anomaly detection techniques to identify the outliers \cite{WXiong5}, or robustify the estimator with other cost functions less sensitive to the deviating samples \cite{WXiong3}.

\begin{figure}[!t]
	\centering
	\includegraphics[width=3.2in]{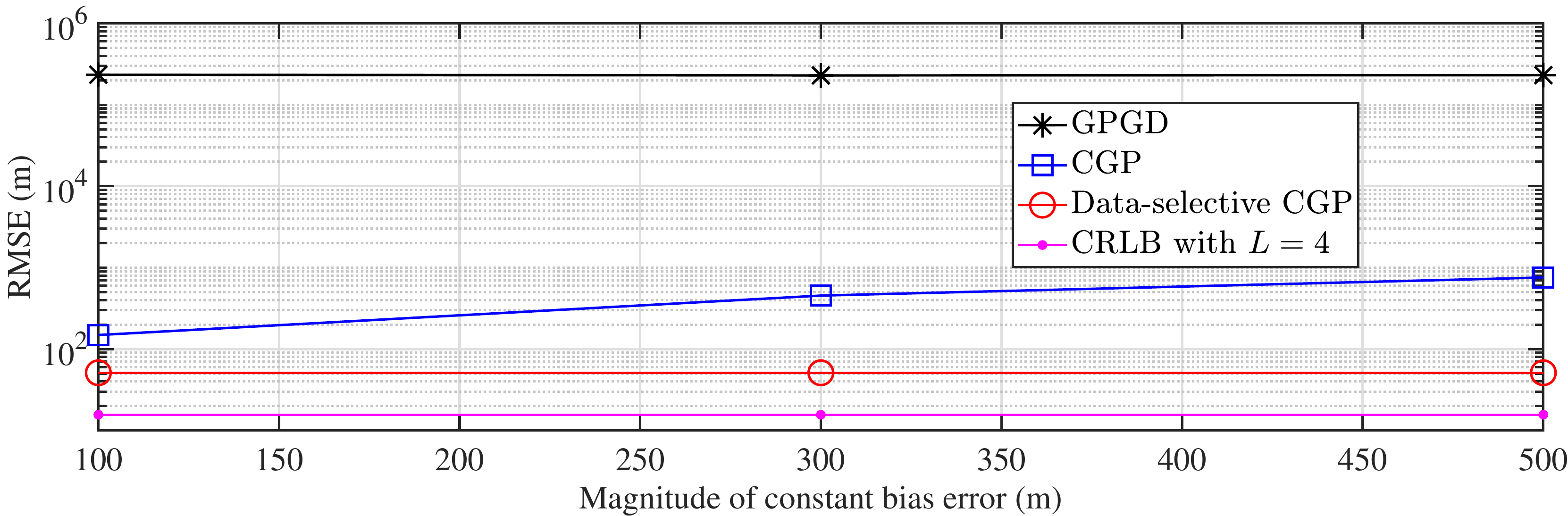}
	\caption{RMSE versus magnitude of constant bias error.} 
	\label{RMSEvsmagn}
\end{figure}

Let us take the data selection strategy in \cite{WXiong5} as an illustration, by replacing the least squares (LS) estimator and the residual-based loss in \cite{WXiong5} with CGP and a new cost function built upon (\ref{ML_est_ref}), respectively. Based on 100 MC trials for each noise level, Fig. \ref{RMSEvsmagn} plots the RMSE of GPGD, CGP, and the data-selective CGP versus the magnitude of a constant bias error added to the RD measurement associated with the fifth sensor. The related experimental and algorithmic parameters in this scenario are set as $\sigma_{1} =...= \sigma_{5} = 10$ m, $N_{\textup{Ptcl}} = 3$, and $N_{\max} = 5$. The CRLB computed using only the first four sensors is shown as well. We see that among three estimators, the data-selective CGP has the minimum RMSE and is closest to the CRLB with four sensors, implying the successful exclusion of the biased sensor observations.

Though the data selection scheme transferred from \cite{WXiong5} has proven to be a workable solution, such a simple workaround is computationally flawed owing to its combinatorial essence. Moreover, the achievable performance of \cite{WXiong5} hinges on the prior knowledge of the number of outliers, not to mention the concerns about localizability raised when discarding the suspicious data. For these reasons, statistically robustifying the location estimator with the help of outlier-resistant loss functions \cite{WXiong3} could be a better choice. As considerable modifications of the algorithm derivations are expected along this path, the topic deserves more systematic analysis in future studies.

\begin{figure}[!t]
	\centering
	\includegraphics[width=3.2in]{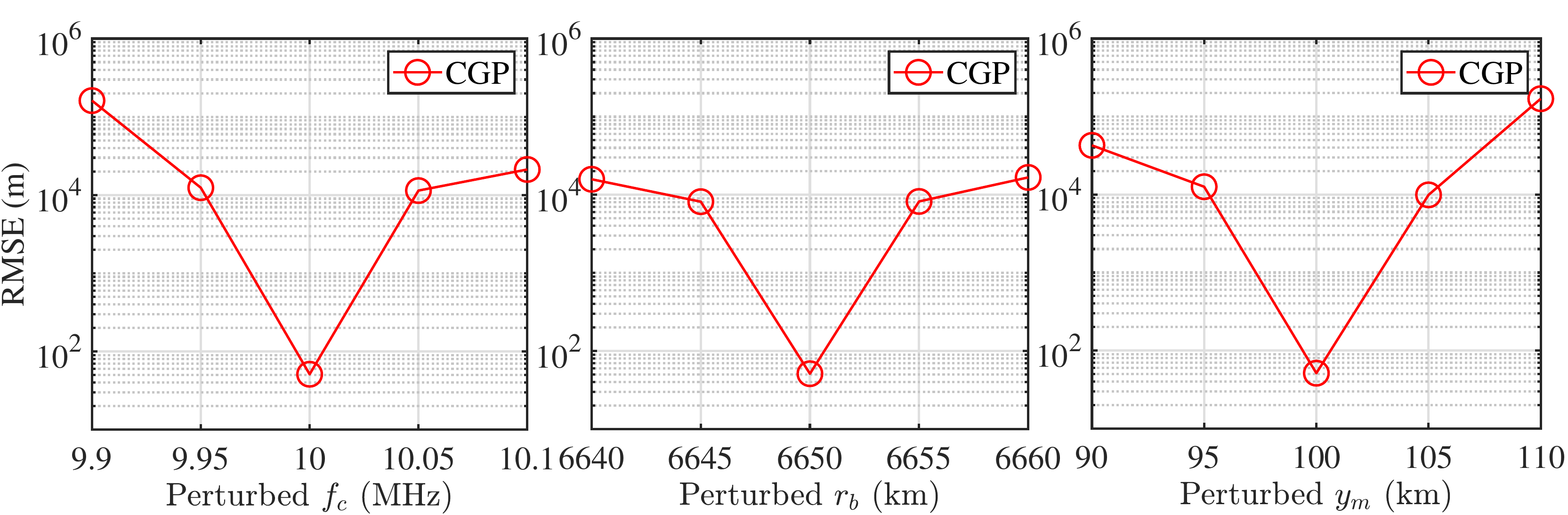}
	\caption{RMSE versus perturbed ionospheric parameters.} 
	\label{RMSEvsPtrb}
\end{figure}

\textit{2) Inaccurate Knowledge of Ionosphere:} The knowledge of the ionospheric profile is often gained from ionograms measured by the ionosonde and exploited later on to evaluate those essential ionospheric parameters \cite{AJain3}. However, several parameters, such as the virtual heights and/or critical frequencies of the ionospheric layers, might not be obtained accurately in practice for various reasons. To showcase their negative effects on the positioning performance of CGP, Fig. \ref{RMSEvsPtrb} plots the RMSE versus $f_c$, $r_b$, and $y_m$ under perturbations. We have for this experiment $L = 4$, $\sigma_1 =...=\sigma_4 = 10$ m, $N_{\textup{Ptcl}} = 3$, and $N_{\max} = 5$. Note that the other two parameters remain unchanged when we alter one of $f_c$, $r_b$, and $y_m$. It is observed that a seemingly slight mismatch between the ionospheric parameters may result in a large localization error. Acquiring knowledge of the ionosphere as accurate as possible or at least keeping the available ionospheric parameters within reasonable bounds is therefore of great significance to HF source geolocation. One of our future plans is to take into consideration the parameter uncertainties when formulating the location estimation problem.

The classical QP model in \cite{TACroft} offers a useful tool for establishing the source-sensor links based on single-hop propagation through a specific layer in the ionosphere. Nevertheless, there is a clear deficiency in the basic QP model: its monolayer assumption does not fit the realities well. Depending on the altitude, the ionosphere filled with ionized atmospheric gases can actually be divided into the D, E, F$_1$, and F$_2$ layers during the day. To address such a deficiency, the more realistic multi-QP (MQP) model \cite{CHou,LYang} will be investigated in the future work.

\textit{3) Limited Computational Resources:} Since CGP involves nested loops, its numerical implementation may sometimes demand significant computational resources. A potential solution for sensor networks with limited computational power is to employ neurodynamic approaches and realize the gradient-type minimizers analogously using designated hardware (e.g., the application-specific integrated circuits), where the computational procedure turns out to be truly distributed and parallel \cite{ZYan,WXiong3}.

\textit{4) Roundness of Earth:} It is known that the spherical approximation is only the simplest model for the shape of Earth in the domain of HF geolocation. More rational options, including the ellipsoidal \cite{LYang} and local approximations, deserve further explorations in the future work.

\section{CONCLUSION AND FUTURE WORK}

\label{Sect_Cc}
This paper presented a globally optimized CGP framework for HF source localization using TDOA measurements under the QP-based ionosphere model, for the purpose of overcoming the initialization problem of an existing gradient-type algorithm. The CGP exploits multiple GP models, each of which is provable to converge to a critical point of the simplified ML formulation. Since a critical point solution in general does not guarantee the successful localization, the PSO is applied to coordinate the GP individuals, thereby performing the search for global optimum with almost sure convergence. The performance of the proposed methodology has been comprehensively evaluated by simulations, and the positioning accuracy superiority of CGP over its competitor has been confirmed. Practical issues and limitations in the real-world use of CGP, such as the biased sensor observations, inaccurate knowledge of ionosphere, limited computational resources, and roundness of earth, have also been discussed respectively. Possible future directions include: 1) statistical robustification of the location estimator, 2) new formulation that takes into account the ionospheric parameter uncertainties, 3) MQP modeling of ionosphere, 4) neurodynamic gradient-type solvers, and 5) more rational approximation strategies for the Earth. For a reality on the ground, collaborations with groups able to capture real HF location data are planned for the future. 






\begin{IEEEbiography}[{\includegraphics[width=1in,height=1.25in,clip,keepaspectratio]{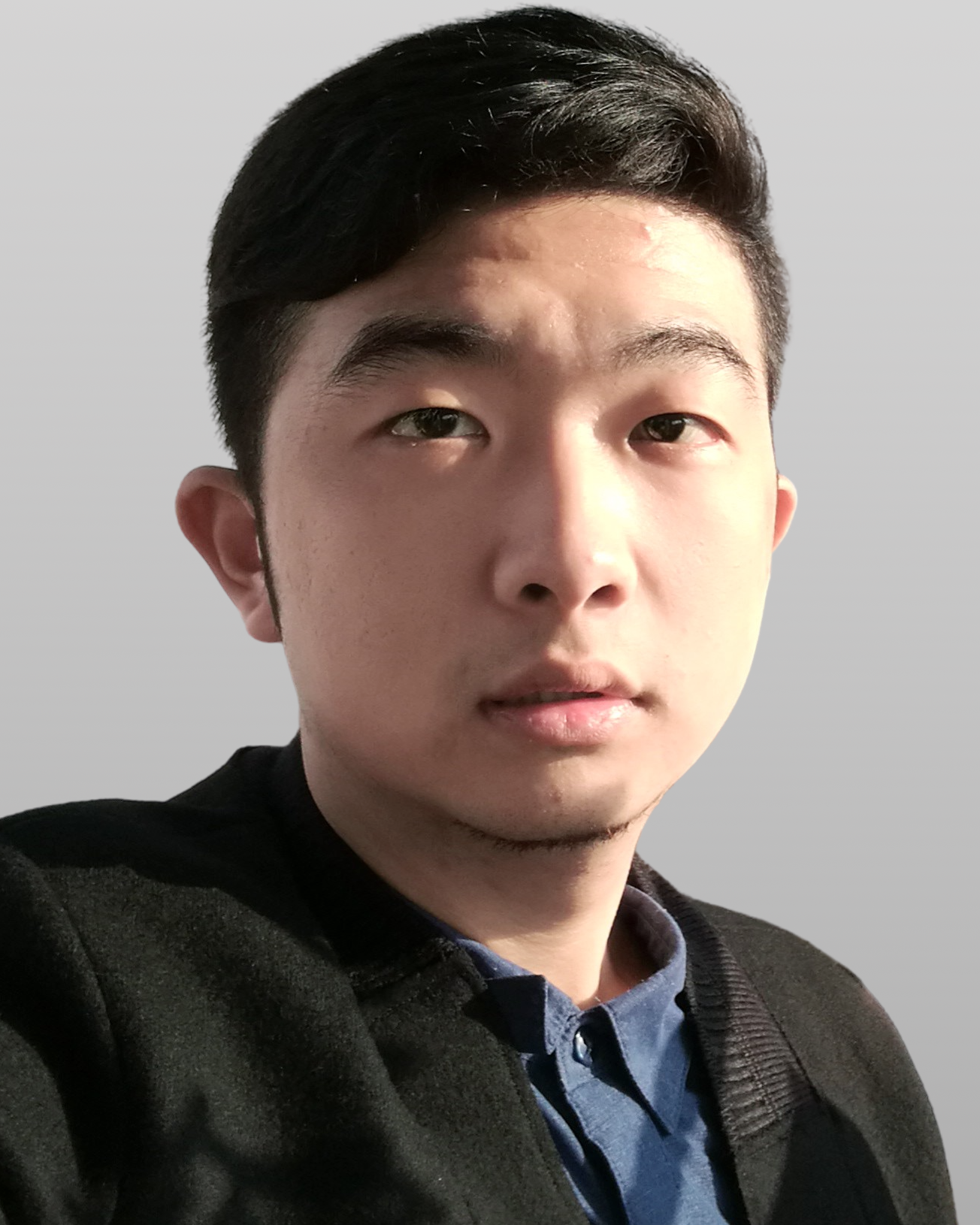}}]{Wenxin~Xiong}{\space}(Graduate Student Member, IEEE) was raised in Wuhan, China. He received the B.Eng. degree in electrical engineering and automation from Northwestern Polytechnical University, Xi'an, China, in 2017, and the M.Sc. degree in electronic information engineering from the City University of Hong Kong, HKSAR, China, in 2018. He is currently pursuing the Dr. rer. nat. degree with the University of Freiburg, Freiburg im Breisgau, Germany, under the supervision of Prof. C. Schindelhauer, Chair of Computer Networks and Telematics (CoNe). His research interests lie broadly in the fields of localization, robust estimation, optimization, and deep learning.
    	
From October 2018 to June 2019, he was a Research Assistant with the Department of Electrical Engineering, City University of Hong Kong. From July 2019 to September 2020, He worked as a Scientific Staff at CoNe.
    
Mr. Xiong serves as a Reviewer for various international journals, including but not limited to \textsc{IEEE Transactions on Information Theory}, \textit{Signal Processing}, \textsc{IEEE Signal Processing Letters}, and \textsc{IEEE Communications Letters}.
\end{IEEEbiography}

\begin{IEEEbiography}[{\includegraphics[width=1in,height=1.25in,clip,keepaspectratio]{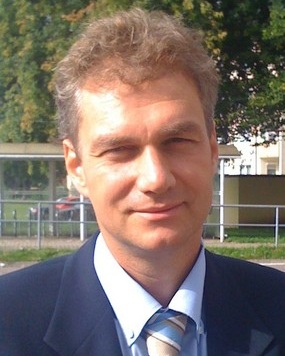}}]{Christian Schindelhauer}
    (Member, IEEE) received the Diploma degree in computer science from Technische Universit\"at Darmstadt, Darmstadt, Germany, in 1992, the Dr.\,rer.\,nat.{} degree from the University of L\"ubeck, L\"ubeck, Germany, in 1996, and the Habilitation degree from the University of Paderborn, Paderborn, Germany, in 2002.
    		
    He was a Scientific Staff Member under the lead of Prof. R. Reischuk from 1992 to 1999 and a Post-Doctoral Researcher with the Theory Group, International Computer Science Institute, Berkeley, CA, USA, under the Group Leader Professor R.~Karp, in 1999. He became a Lecturer with the University of Paderborn in 2002, after being a Post-Doctoral Researcher since 2001. Since 2006, he has been a Professor of Computer Networks and Telematics with the University of Freiburg, Freiburg im Breisgau, Germany. He has authored more than 150 peer-reviewed conference papers and journal articles. His current research interests include distributed algorithms, peer-to-peer networks, mobile ad hoc networks, wireless sensor networks, localization algorithms, storage networks, and coding theory.
    	
    Christian~Schindelhauer has served as a program committee member of over 50 conferences. He received the Prof.-Otto-Roth-Award from the Universit\"at L\"ubeck for an outstanding Ph.D.{} thesis in 1997. He is a co-editor of the \textit{Telecommunication Systems} journal and a member of the Association for Computing Machinery and Gesellschaft der Informatik.
\end{IEEEbiography}

\begin{IEEEbiography}[{\includegraphics[width=1in,height=1.25in,clip,keepaspectratio]{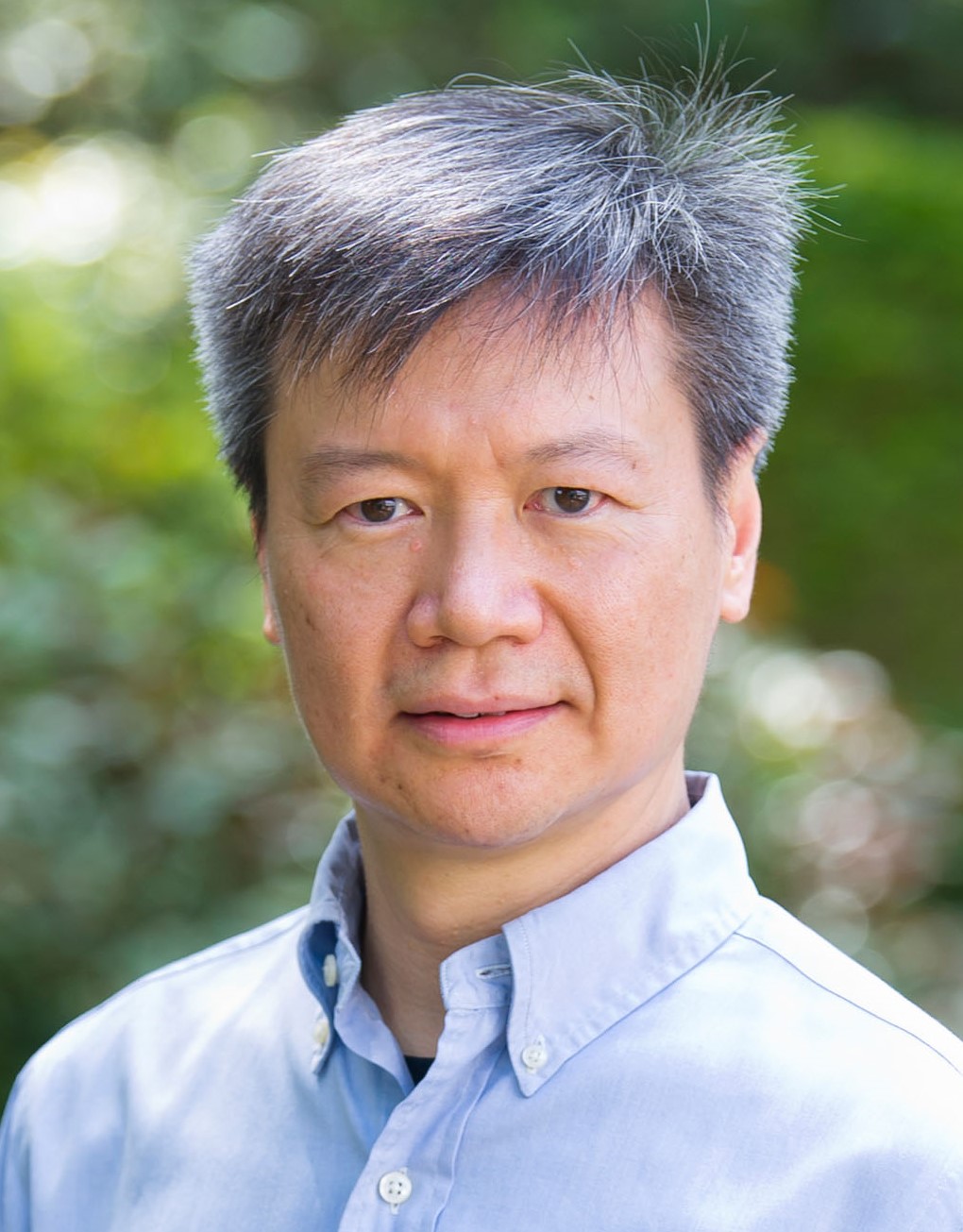}}]{Hing~Cheung~So}
	(Fellow, IEEE) was born in Hong Kong. He received the B.Eng. degree from the City University of Hong Kong and the Ph.D. degree from The Chinese University of Hong Kong, both in electronic engineering, in 1990 and 1995, respectively. From 1990 to 1991, he was an Electronic Engineer with the Research and Development Division, Everex Systems Engineering Ltd., Hong Kong. During 1995–1996, he was a Postdoctoral Fellow with The Chinese University of Hong Kong. From 1996 to 1999, he was a Research Assistant Professor with the Department of Electronic Engineering, City University of Hong Kong, where he is currently a Professor. His research interests include detection and estimation, fast and adaptive algorithms, multidimensional harmonic retrieval, robust signal processing, source localization, and sparse approximation.
			
	He has been on the editorial boards of \textsc{IEEE Signal Processing Magazine} (2014--2017), \textsc{IEEE Transactions on Signal Processing} (2010--2014), \textit{Signal Processing} (2010--), and \textit{Digital Signal Processing} (2011--). He was also the Lead Guest Editor for the \textsc{IEEE Journal of Selected Topics in Signal Processing}, special issue on ``Advances in Time/Frequency Modulated Array Signal Processing'' in 2017. In addition, he was an Elected Member in Signal Processing Theory and Methods Technical Committee (2011--2016) of the IEEE Signal Processing Society where he was the Chair in the awards subcommittee (2015--2016).
\end{IEEEbiography}


\begin{thebibliography}{[31d]}  

	

	\bibitem{AJain1}
	A. Jain, P. Pagani, R. Fleury, M. M. Ney, and P. Pajusco, ``Efficient time domain HF geolocation using multiple distributed receivers,'' in \textit{Proc. IEEE 11th Euro. Conf. Antennas Propag.}, Paris, France, Mar. 2017, pp. 1852--1856.


	\bibitem{TWang}
	T. Wang, X. Hong, W. Liu, A. M.-C. So, and K. Yang, ``Geolocation of unknown emitters using TDOA of path rays through the ionosphere by multiple coordinated distant receivers,'' in \textit{Proc. IEEE Int. Conf. Acoust., Speech Signal Process.}, Calgary, AB, Canada, Sep. 2018, pp. 3509--3513.


	\bibitem{SHuang}
	S. Huang, Y. -M. Pun, A. M. -C. So, and K. Yang, ``A provably convergent projected gradient-type algorithm for TDOA-based geolocation under the quasi-parabolic ionosphere model,'' \textit{IEEE Signal Process. Lett.}, vol. 27, pp. 1335--1339, 2020.


	\bibitem{AJain2}
	A. Jain, P. Pagani, R. Fleury, M. M. Ney, and P. Pajusco, ``Accurate time difference of arrival estimation for HF radio broadcast signals,'' \textit{Microw. Opt. Techn. Lett.}, vol. 60, no. 6, pp. 1406--1410, Jun. 2018.


	\bibitem{AJain3}
	A. Jain, P. Pagani, R. Fleury, M. M. Ney, and P. Pajusco, ``HF source geolocation using an operational TDoA receiver network: Experimental results,'' \textit{IEEE Antennas Wireless Propag. Lett.}, vol. 17, no. 9, pp. 1643--1647, Sep. 2018.


	\bibitem{DBilitza}
	D. Bilitza, ``International reference ionosphere 2000,'' \textit{Radio Sci.}, vol. 36, no. 2, pp. 261--275, 2001.


	\bibitem{AHDeVoogt}
	A. H. De Voogt, ``The calculation of the path of a radio-ray in a given ionosphere,'' \textit{Proc. IRE}, vol. 41, no. 9, pp. 1183--1186, Sep. 1953.



	\bibitem{TACroft}
	T. A. Croft and H. Hoogasian, ``Exact ray calculations in a quasi-parabolic ionosphere with no magnetic field,'' \textit{Radio Sci.}, vol. 3, pp. 69--74, 1968.


	\bibitem{CHou}
	C. Hou, G. Ke, and Y. Fu, ``The sky-wave radar detection performance computing based on the dynamic ionospheric model,'' \textit{Neurocomputing}, vol. 151, Part 3, pp. 1305--1315, Mar. 2015.


	\bibitem{EGBakhoum}
	E. G. Bakhoum, ``Closed-form solution of hyperbolic geolocation equations'' \textit{IEEE Trans. Aerosp. Electron. Syst.}, vol. 42, no. 4, pp. 1396--1404, Oct. 2006.


	\bibitem{SHuangCode}
	S. Huang, ``Fast and accurate algorithms for geolocation and multiple source localization,'' Ph.D. dissertation, Chinese Univ. of Hong Kong, Hong Kong, China, 2021. Source Code Available: \url{https://github.com/samwong1993}




	\bibitem{ZYan}
	Z. Yan, J. Fan, and J. Wang, ``A collective neurodynamic approach to constrained global optimization,'' \textit{IEEE Trans. Neural Netw. Learn. Syst.}, vol. 28, no. 5, pp. 1206--1215, May 2017.




	\bibitem{PSAndrews}
	P. S. Andrews, ``An investigation into mutation operators for particle swarm optimization,'' in \textit{Proc. 2006 IEEE Int. Conf. Evol. Comput.}, Vancouver, BC, Canada, Jul. 2006, pp. 1044--1051.



	\bibitem{GSSales}
	G. S. Sales, ``High frequency (HF) radiowave propagation,'' Phillips Laboratory, Sci. Rep. \#6, PL-TR-92-2123, 1992.



	\bibitem{AJain4}
	A. Jain, P. Pagani, R. Fleury, M. M. Ney, and P. Pajusco, ``Cross-channel sounding for HF geolocation: Concepts and experimental results,'' in \textit{Proc. 12th Eur. Conf. Antennas Propag.}, London, UK, Apr. 2018, pp. 1--5.



	\bibitem{NXia}
	N. Xia, W. Wei, J. Li, and X. Zhang, ``Kalman particle filtering algorithm for symmetric alpha-stable distribution signals with application to high frequency time difference of arrival geolocation,'' \textit{IET Signal Process.}, vol. 10, no. 6, pp. 619--625, Aug. 2016.


	\bibitem{NXia2}
	N. Xia and B. Xing, ``A direct localization method for HF source geolocation and experimental results,'' \textit{IEEE Antennas Wireless Propag. Lett.}, vol. 20, no. 5, pp. 728--732, May 2021.


	\bibitem{FJavidrad}
	F. Javidrad and M. Nazari, ``A new hybrid particle swarm and simulated annealing stochastic optimization method,'' \textit{Appl. Soft Comput.}, vol. 60, pp. 634--654, Nov. 2017.



	\bibitem{SCEsquivel}
	S. C. Esquivel and C. A. Coello Coello, ``On the use of particle swarm optimization with multimodal functions,'' in \textit{Proc. 2003 Congr. Evol. Comput.}, vol. 2, Canberra, Australia, Dec. 2003, pp. 1130--1136.



	\bibitem{UMAscher}
	U. M. Ascher and C. Greif. \textit{A First Course in Numerical Methods.} Philadelphia, PA, USA: SIAM, 2011.






	\bibitem{HAttouch}
	H. Attouch, J. Bolte, and B. F. Svaiter, ``Convergence of descent methods for semi-algebraic and tame problems: Proximal algorithms, forward-backward splitting, and regularized Gauss-Seidel methods,'' \textit{Math. Program.}, vol. 137, no. 1--2, pp. 91--129, 2013.


	\bibitem{ASLewis}
	A. S. Lewis, D. R. Luke, and J. Malick, ``Local linear convergence for alternating and averaged nonconvex projections,'' \textit{Found. Comput. Math.}, vol. 9, no. 4, pp. 485--513, Aug. 2009.


	\bibitem{MFrank}
	M. Frank and P. Wolfe, ``An algorithm for quadratic programming,'' \textit{Nav. Res. Logistics Quart.}, vol. 3, no. 1--2, pp. 95–110, 1956.


	\bibitem{CHildreth}
	C. Hildreth, ``A quadratic programming procedure,'' \textit{Naval Res. Logist. Quart.}, vol. 4, pp. 79--85, 1957.



	\bibitem{FJSolis}
	F. Solis and R. Wets, ``Minimization by random search techniques,'' \textit{Math. Oper. Res.}, vol. 6, no. 1, pp. 19--30, 1981.



	\bibitem{GoogleEarth}
	Google Earth Help, ``Find \& use location coordinates,'' 2021. [Online]. Available: \url{https://support.google.com/earth/answer/148068}




	\bibitem{WXiong3}
	W. Xiong \textit{et al.}, ``TDOA-based localization with NLOS mitigation via robust model transformation and neurodynamic optimization,'' \textit{Signal Process.}, vol. 178, 107774, Jan. 2021.


	\bibitem{WXiong5}
	W. Xiong \textit{et al.}, ``Data-selective least squares methods for elliptic localization with NLOS mitigation,'' \textit{IEEE Sensors Lett.}, vol. 5, no. 7, pp. 1--4, Jul. 2021.



	\bibitem{LYang}
	L. Yang, H. Gao, Y. Ling, and B. Li, ``Localization method of wide-area distribution multistatic sky-wave over-the-horizon radar,'' \textit{IEEE Geosci. Remote Sens. Lett.,}, vol. 19, pp. 1-5, 2022, Art. no. 3500305.


	

\end{thebibliography}
\end{document}